%% file: main.tex
\documentclass[sigconf]{acmart}

\setcopyright{rightsretained}

\acmConference[]{}{}{} 
\acmISBN{}
\acmDOI{}

\usepackage[utf8]{inputenc}
\usepackage[T1]{fontenc}

\graphicspath{{figs/}}
\usepackage{color}
\usepackage[tight]{subfigure}
\usepackage{booktabs}
\usepackage{multirow}
\usepackage{enumitem}
\usepackage[english]{babel}
\usepackage{import}
\usepackage{tikz}
\usetikzlibrary{decorations.pathmorphing,decorations.pathreplacing,shadows,shapes}

\usepackage{url}

\usepackage{balance}

\usepackage{array}
\newcolumntype{L}[1]{>{\raggedright\let\newline\\\arraybackslash\hspace{0pt}}m{#1}}
\newcolumntype{C}[1]{>{\centering\let\newline\\\arraybackslash\hspace{0pt}}m{#1}}
\newcolumntype{R}[1]{>{\raggedleft\let\newline\\\arraybackslash\hspace{0pt}}m{#1}}

\input{commands}

\title{Connecting the World of Embedded Mobiles: The RIOT Approach to Ubiquitous Networking for the IoT}
\author{Martine Lenders}
\affiliation{%
  \institution{Freie Universit\"at Berlin}
}
\email{m.lenders@fu-berlin.de}

\author{Peter Kietzmann}
\affiliation{%
  \institution{HAW Hamburg}
}
 \email{peter.kietzmann@haw-hamburg.de}

 \author{Oliver Hahm}
\affiliation{%
  \institution{RIOT}
}
 \email{oleg@riot-os.org}

\author{Hauke Petersen}
\affiliation{%
  \institution{Freie Universit\"at Berlin}
}
\email{hauke.petersen@fu-berlin.de}

\author{Cenk G{\"u}ndo\u{g}an}
\affiliation{%
  \institution{HAW Hamburg}
}
 \email{cenk.guendogan@haw-hamburg.de}

\author{Emmanuel Baccelli}
\affiliation{%
\institution{Inria}
}
 \email{emmanuel.baccelli@inria.fr}

 \author{Kaspar Schleiser}
\affiliation{%
  \institution{RIOT}
}
 \email{kaspar@schleiser.de}

\author{Thomas C. Schmidt}
\affiliation{%
  \institution{HAW Hamburg}
}
 \email{t.schmidt@haw-hamburg.de}

\author{ Matthias W{\"a}hlisch}
\affiliation{%
  \institution{Freie Universit\"at Berlin}
}
\email{m.waehlisch@fu-berlin.de}

\begin{document}

%%%%%

% Use the following at camera-ready time to suppress page numbers.
% Comment it out when you first submit the paper for review.
%\thispagestyle{empty}

%\subsection*{Abstract}
\begin{abstract}
\input{tex/abstract}

\end{abstract}

\maketitle
\renewcommand{\shortauthors}{Lenders et al.}

%\maketitle

\input{tex/introduction}

\input{tex/challenges-and-related-work}

\input{tex/network-stacks}

\input{tex/gnrc}
\input{tex/evaluation}

\input{tex/conclusion}

\subsection*{Availability of Source Code}
We will make our source code publicly available on GitHub.
This will not only include the RIOT networking subsystem and stacks but also all measurement applications, to improve both IoT deployment and reproducibility of our experiments.

\iffalse
\bigskip
%\paragraph{Acknowledgements}
\section*{Acknowledgements}
%Removed due to double-blind submission.
%\iffalse
Thanks to the RIOT community including special thanks to Thomas Eichinger, Johann Fischer, Nick van IJzendoorn, Cenk G\"undogan, Raphael Hiesgen, Ren\'e Kijewski, Martin Landsmann, Ian Martin, Christian Mehlis, Joakim Nohlgard, Ludwig Ortmann, Kevin Roussel, Lotte Steenbrink, Hinnerk Van Bruinehsen, Takuo Yonezawa.
The authors also wish to acknowledge (in alphabetical order) Francisco Acosta, C\'edric Adjih, Carsten Bormann, Maike Gilliot, Vlado Handziski, Pekka Nikander, Jochen Schiller, Thiemo Voigt, and Tianyin Xu for useful contributions, feedback, and suggestions.
Special thanks to Heiko Will for having been under fire.
Parts of this work has been sponsored by French Agence Nationale de la Recherche and German Federal Ministry of Education and Research.
We also would like to thank the founding institutions of RIOT, Freie Universit\"at Berlin, INRIA, and Hamburg University of Applied Sciences.
Finally, we would like to thank Peter Schmerzl for his inspiring ideas on operating systems, which he has implemented in SchmerzlOS.
\fi

\iffalse
\section{Availability}

RIOT code is open source, and publicly available on GitHub at

\begin{center}
{\tt https://github.com/RIOT-OS/RIOT}
\end{center}
\fi

\balance
{\footnotesize \bibliographystyle{abbrv}
\bibliography{rfcs,RIOT-paper}}

%\theendnotes

\end{document}

%% file: commands.tex
\usepackage{pifont}
\newcommand{\cmark}{\ding{51}}%
\newcommand{\xmark}{\ding{56}}%
\newcommand{\umark}{\ding{114}}

\usepackage{xspace}

\newcommand{\one}{({\em i})\xspace}
\newcommand{\two}{({\em ii})\xspace}
\newcommand{\three}{({\em iii})\xspace}
\newcommand{\four}{({\em iv})\xspace}
\newcommand{\five}{({\em v})\xspace}

\makeatletter
\renewcommand{\paragraph}[1]{\vspace*{0.03in}\noindent{\bf #1.}\hspace{0.25ex \@plus1ex \@minus.2ex}}
\makeatother

%% file: tex/abstract.tex
The Internet of Things (IoT) is rapidly evolving based on low-power compliant protocol standards that extend the Internet into the embedded world. 
Pioneering implementations have proven it is feasible to inter-network very constrained devices, but had to rely on peculiar cross-layered designs and offer a minimalistic set of features.
In the long run, however, professional use and massive deployment of IoT devices require  full-featured, cleanly composed, and flexible network stacks.

This paper introduces the networking architecture that turns  RIOT into a powerful IoT system, to enable low-power wireless scenarios.
RIOT networking offers (i) a modular architecture with generic interfaces for plugging in drivers, protocols, or entire stacks, (ii) support for multiple heterogeneous interfaces and stacks that can concurrently operate, and (iii) GNRC, its cleanly layered, recursively composed default network stack. 
We contribute an in-depth analysis of the communication performance and resource efficiency of RIOT, both on a micro-benchmarking level as well as by comparing IoT communication across different platforms.
Our findings show that, though it is based on significantly different design trade-offs, the networking subsystem of RIOT achieves a performance equivalent to that of Contiki and TinyOS, the two operating systems which pioneered IoT software platforms.

%% file: tex/introduction.tex
\section{Introduction}

The Internet is constantly evolving. While Internet mobility was proposed and developed 15 years ago \cite{RFC-3344,RFC-3775}, followed by a decade of massive deployment for Internet connected mobiles, we are now facing a strong momentum towards inter-networking the world of small embedded devices. It is the vision of the Internet of Things (IoT) \cite{atzori2010internet} to connect the physical world by means of sensors and actuators, thereby enabling entirely new services and drastic efficiency improvements for industry, consumers, and the common public environment. 
%The IoT is expected to fuel a trillion dollar business selling the idea of unhindered harvesting of immersive data.

The initial mobile revolution was grounded on smartphones and handhelds, smartTVs and other single-board computers such as the RasberryPi. These machines have enough resources to run common operating systems such iOS, Linux, BSD, Windows, or their derivatives (e.g., OpenWRT, Android, or uClinux), and use traditional network stacks to run IP protocols over Wi-Fi or Ethernet.

The embedded, low-end IoT~\cite{RFC-7228} consists of sensors, actuators, and various ultra-cheap communication modules, which have orders of magnitude less resources: kBytes of memory, MHz CPU speed, mW of power, and these limitations are expected to last \cite{waldrop:2016,Quartz-IoT-Moore}. 
Due to the lack of resources, low-end IoT devices cannot run conventional operating systems, and instead use IoT operating systems \cite{hahm2015survey} such as Contiki \cite{Dunkels+:2004} or RIOT \cite{hahm+:2013}. 
Equally important from the network perspective, low-end IoT devices typically communicate via low-power radio (e.g., IEEE 802.15.4, Bluetooth LE, lpWAN).
Compared to WiFi or Ethernet, low-power radios impose a variety of constraints which led to the standardization of new protocol elements. 
Perhaps the most prominent example is the 6LoWPAN stack \cite{sheng2013survey}, based on IPv6 and an adaptation layer to low power lossy networks, while retaining interoperability with the core  Internet.

In this paper, we  take a system perspective on small, embedded devices and discuss viable options and limitations of how to turn them into first class Internet citizens. We report on the design and integration of a  high-level networking architecture including a full-featured, layered network stack in RIOT that remains competitive with monolithically optimized approaches. In detail,  the contributions of this paper read:

\begin{enumerate}
	\item We present an IoT subsystem built of clean layers around well-abstracted APIs that enable full-featured networking  for all available drivers including IEEE 802.15.4, Ethernet, LoRa, CAN. Conversely, drivers or upper protocol layers can be seamlessly augmented by new networking functions such as ICN or SDN.

	\item This networking subsystem is shown to simultaneously operate on multiple, heterogeneous transceivers. Hence it enables technology handovers and routing between varying access links.

  \item We present GNRC, the default network stack in RIOT, which takes full advantage of the multi-threading model supported by RIOT to foster a clean protocol separation via well-defined interfaces and IPC.
  %We show that significant overhead is saved by using thread-based message queues in embedded systems, in contrast to centralized queues.

  \item We evaluate GNRC on real hardware and compare it with lwIP and emb6, two prominent IoT network stacks.
  We show that GNRC provides comparable performance, while allowing higher configurability and easier updates of parts of the stack.
  Our evaluation provides a benchmark for non-trivial scenarios with multi-threading on low-end IoT devices.

  \item We compare GNRC with a network stack for a future, information-centric network (ICN), both implemented in our proposed networking subsystem.
  We identify performance aspects that inherently follow from the chosen network architecture.
    %\todoEB{This sounds like a more profound subject, can it seriously be brushed over in this paper?}

  \item We make our source code publicly available, to improve IoT deployment and reproducibility of
our experiments.
\end{enumerate}

The remainder of this paper is structured as follows. In Section~\ref{sec:background}, we present background, requirements, and related work for IoT network stacks.
We introduce the RIOT networking subsystem and its default network stack as well as 3rd party stacks, in Section~\ref{sec:nw-stack} and Section~\ref{sec:gnrc}  respectively.
In Section~\ref{sec:neteval}, we present our performance evaluation.
We conclude in Section~\ref{sec:conclusions}.

%% file: tex/challenges-and-related-work.tex
\section{The Need for a Full-featured Embedded Internet, Reviewing Related Work}
\label{sec:background}

As the complexity of software for embedded devices has increased over the last decade, it has become state-of-the-art to use operating systems even on memory and CPU constrained machines. The network subsystem is a key component for building the Internet of Things. 
%It provides interfaces to operate specific stacks between the network device drivers and the applications.
%A full-featured network stack is one of the most complex pieces of software to run on an embedded platform.
A \emph{full-featured stack} allows for a complete implementation of all protocol specifications per design.

 Throughout this paper, we assume that the network stack is built atop an OS that provides the following features: \one support of threads/processes, \two a lightweight process model, \three efficient inter-process communication (IPC), \four a lean common hardware abstraction, and \five a memory foot-print suitable for IoT devices. Assumptions \one-\three allow for a modular network stack that is split over multiple processes without a significant overhead through administrative data structures. Assumption \four enables the network stack to be independent from specific IoT hardware platforms and network devices. 
%We also assume that the OS allows the network stack to be open source, maintained by a lively community (similarly to Linux).
An operating system such as RIOT \cite{hahm+:2013} implements these basic system assumptions.

It is worth noting that the BSD~4.4 network stack~\cite{chesson1975} and its successors in Unix and Linux systems \cite{wehrle2004} followed fundamentally different design directives, focusing predominantly on throughput as apparent from the way buffers are designed.
%were originally developed in times when the memory constraints of a typical computer were roughly comparable with that of current IoT devices.
%However, their development followed fundamentally different design objectives, focusing predominantly on throughput (this manifests itself e.g., in the way buffers are designed).
During decades of development this led to drastically increased system resources and worked in a direction opposite to low-end IoT devices. Hence, we had to reconsider basic principles and to draw a design from scratch.

We will now present a first reality check and then discuss design choices of previous work for implementing specific features.

\subsection{A First Glimpse into the Small World}
\label{sec:showcase}

\begin{figure}
    \centering
    \resizebox{0.99\columnwidth}{!}{\input{figs/multi-interface}}
    \caption[]{Sensing and forwarding---a deployment use case of heterogeneous IoT networking.}
    \label{fig:multi-interface}
\end{figure}
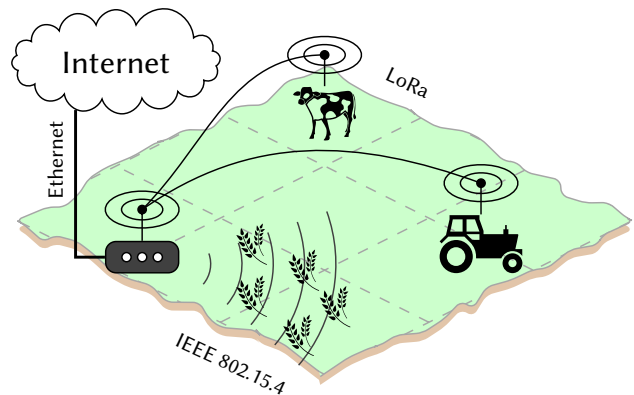

A typical IoT use case is displayed in Figure~\ref{fig:multi-interface}. 
It consists of sensors attached to elements in the field -- a tractor and a cow -- as well as a data collection and aggregation system. 
Embedded nodes are deployed as both, the wireless field sensors and the data collectors. 
The latter may be connected to some (wired) infrastructure. 
As shown in Figure \ref{fig:multi-interface}, the sample deployment consists of two RIOT sensors, one equipped with LoRa and one with 802.15.4 LoWPAN wireless interface, as well as one RIOT router providing both radio links and a wired Ethernet uplink. 
In a lab experiment, we demonstrate this with both sensor nodes running continuous packet streams of single sensor values, and a RIOT border router which forwards data upstream. All nodes are very constrained iotlab-m3 boards (ARM Cortex M3, 32-bits, 72 Mhz, 64kB RAM). 

This sample deployment operates very reliably at a rate up to 5 packets per second (packet loss $ \ll 1$ \%) per sensor. When increasing the data rate to 10 packets per second or beyond, the LoRa link turns into a noisy service of quickly increasing packet loss. This is due to its inability of transmitting data at the requested rates. On the contrary, the 802.15.4 LowPAN layer continues stable operations and so does the RIOT networking and forwarding system. We can thus conclude that the networking subsystem introduced in this paper can successfully serve the needs the IoT target scenarios in the presence of  transmission layers characteristic for highly constrained environments.

For the next reality, check we want to shed light on the  performance of our  full-featured network stack following a cleanly layered design in comparison to other prominent, but cross-layer optimized solutions.   

For this, we chose ultra-constrained devices, the Zolertia Z1 nodes (msp430 16-bit MCU, 16 MHz, 92 KB Flash, 8 KB RAM, on-board 2,4 GHz 802.15.4 radio TI cc2420) that have joint support of the two pioneering embedded operating systems Contiki and TinyOS, as well as RIOT. Each OS is deployed with its native stack---GNRC for RIOT as contributed in this work. We implemented a simple test application that periodically transmits a UDP packet of 20 Bytes payload and measured the transmission times over the air, differentiating with respect to stack and driver processing times at both sides  as well as the wireless transmission (incl. hardware processing).

\begin{figure}
  \centering
  \includegraphics[width=1.0\columnwidth]{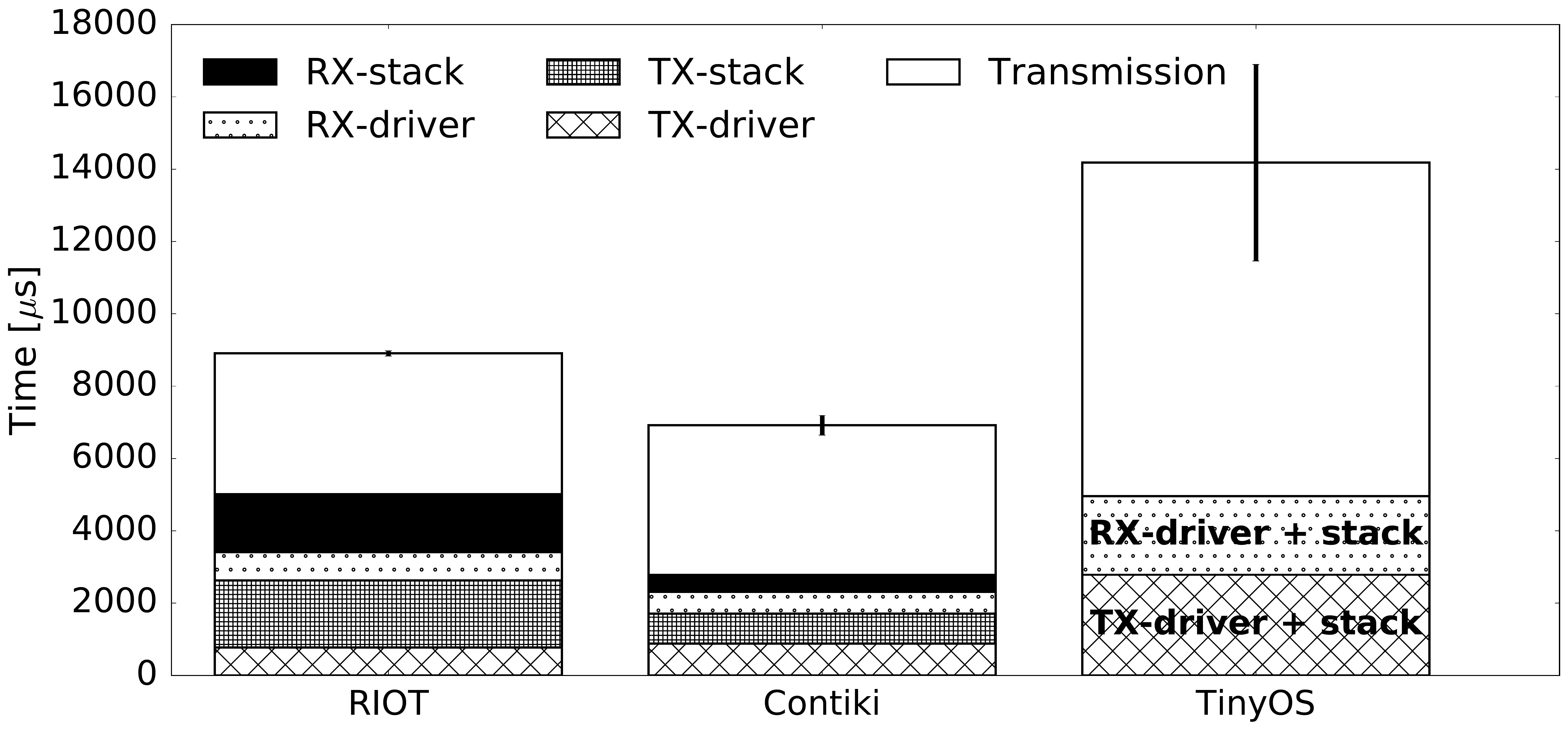}
  \caption{Communication times between tiny IoT devices}
  \label{fig:z1-contiki-riot-tiny}
\end{figure}

Results for a single packet transmission are displayed in Figure \ref{fig:z1-contiki-riot-tiny} and clearly encourage. Even though of quite distinct design, all stacks perform on the same scale with fluctuating diversity. While RIOT is a bit slower in stack and driver processing, the actual transmission over IoT radios stands out as the dominating part. Even the large fluctuations in wireless transmissions could compensate for the overheads obtained from software complexity.   

While we will argue along the lines of this article that the RIOT networking subsystem follows a strictly modular design with useful, high-level abstractions, we can proceed in the understanding that  this does not sacrifice performance in the small world of IoT devices.

\subsection{Required Functionalities in the IoT}

\paragraph{Support for Multiple Network Interfaces}
%IoT scenarios do not only include basic sensors with a microcontroller and a single low-power radio, but also 
%devices which have multiple radio interfaces (e.g., IEEE 802.15.4 and Bluetooth) to improve reliability or improve throughput.
%Thus, the network stack must be able to handle multiple network interfaces.
Currently, only some stacks \cite{dunkels2001lwip,borchert2012ciao,hutchinson1988design,bhatti2005mantis} support multiple network interfaces in parallel.
To reduce memory overhead, most implementations~\cite{han2005dynamic,hui2008ip,watteyne+:2012,dunkels2007rime} are based on a single scalar data structure instead of an array or linked-list.
This design choice limits application scenarios.
It may allow for border router scenarios using serial line protocols (SLIP, PPP) to bridge 802.15.4 and Ethernet,
such as in BLIP, uIP, and Rime, but will not allow for fully flexible IoT devices.
We show that---if designed carefully---the overhead of multi-interface support is negligible compared to single interface support, even on constrained devices.

\paragraph{Parallel Data Handling}
Some embedded network stacks achieve their small memory footprint by reducing their functionality to the point where they are only able to handle a single network packet at a time.
While this might be reasonable in some use cases, this is unrealistic in general.
Using reactive protocols is common, in particular for dynamic neighbor discovery.
%IPv6 over spontaneous wireless networking, multiple services run in parallel and use .
%both routing and neighbor discovery protocols are tightly coupled to data transfers between nodes.
%For example, both routing and neighbor discovery protocols are tightly coupled to data transfers between nodes.
Thus, the network stack must be able to buffer multiple packets, usually implemented by advanced data structures.

uIP and Rime use a single byte~array that only stores a single packet, inherently preventing parallel data handling.
Arrays of arrays or ring buffers of data pointers allow buffering of multiple packets, such as in OpenWSN, Mantis~OS, and CiAO/IP.
However, those data structures store packet data sequentially and thus conflict with dynamic parsing of headers.

More appealing are abstract data structures where a packet is dynamically stored in a pre-allocated memory region, a so called \emph{arena}~\cite{hanson1990fast}.
Implementations of an arena organize packet data either using fixed or variable sized chunks.
lwIP uses an arena with fixed-sized chunks, leading to waste of memory.
BLIP and our proposal optimize memory consumption by choosing chunk sizes according to packet lengths.

%packets have similar format ... not a problem with external fragmentation

\begin{table*}%[!t]
  \centering
  \small
    \begin{tabular}{R{.10\linewidth}C{.08\linewidth}C{.08\linewidth}C{.09\linewidth}
                    C{.07\linewidth}C{.07\linewidth}C{.06\linewidth}C{.08\linewidth}
                    C{.08\linewidth}C{.08\linewidth}}
    \toprule
    & \multicolumn{3}{c}{\textbf{Applied Concepts}} &
        \multicolumn{6}{c}{\textbf{Achieved Functionalities}} \\
    \cmidrule(rl){2-4}
    \cmidrule(rl){5-10}
      % & \oone & \otwo & \othree & \oone & \otwo & \othree & \othree \\
      \textbf{Network stack} & Data Sharing & Data Structure NW IF  & Packet Buffer &
        Loosely Coupled Comp. & Clean API & Modular & Multi-NW~IF & Dyn. Header Parsing & Parallel Data Handling \\
    \midrule
      GNRC (RIOT) & message-based & array & arena (variable) &
        \cmark & \cmark & \cmark & \cmark & \cmark & \cmark \\
	    lwIP (Mynewt)~\cite{dunkels2001lwip} & platform-dependent & linked-lists & arena (fixed) &
        \xmark & \xmark & \cmark & \cmark & \cmark & \cmark \\
	    Zephyr~\cite{zephyr2017project-nw} & message-based & array & arena (variable) & 
	 \xmark & \xmark & \cmark & \cmark & \cmark & \cmark \\
      CiAO/IP~\cite{borchert2012ciao} & aspect-oriented & linked-list & ring buffer &
        \xmark & \xmark & \cmark & \cmark & \xmark & \cmark \\
      x-Kernel~\cite{hutchinson1988design} & message-based & unknown & unknown &
        \xmark & \xmark & \cmark & \cmark & \umark & \umark \\
      Mantis OS~\cite{bhatti2005mantis} & message-based & array & multi-array &
        \xmark & \xmark & \xmark & \cmark & \xmark & \cmark \\
      SOS~\cite{han2005dynamic} & message-based & scalar & arena (fixed) &
        \xmark & \xmark & \cmark & \xmark & \xmark & \cmark \\
    % \cmidrule(rl){1-9}
    %   SP/NL~\cite{ee2006modular} & AOP & unknown & unknown &
    %     \cmark \xmark & & \cmark & \umark & \cmark \\
      BLIP (TinyOS)~\cite{hui2008ip} & event-driven & scalar & arena (variable) &
        \xmark & \xmark & \xmark & \xmark & \cmark & \cmark \\
    % \cmidrule(rl){1-9}
    %   da~Silva~Santos~\cite{da2013routing} & unknown & unknown & ring buffer &
    %     \cmark & \xmark & \umark & \umark & \umark \\
      OpenWSN~\cite{watteyne+:2012} & event-driven & scalar & multi-array &
        \xmark & \xmark & \xmark & \xmark & \xmark & \cmark \\
      uIP (Contiki)~\cite{dunkels2002uip} & event-driven & scalar & array &
        \xmark & \xmark & \xmark & \xmark & \xmark & \xmark \\
      Rime (Contiki)~\cite{dunkels2007rime} & event-driven & scalar & array &
        \xmark & \xmark & \xmark & \xmark & \xmark & \xmark \\
    \bottomrule
  \end{tabular}
  \caption{Comparison of related work with the RIOT default network stack, \cmark~supported, \xmark~not supported, \umark~unknown.}
  \label{tab:related-work}
\end{table*}

\paragraph{Horizontal and Vertical Modularity}
The network stack consists of horizontal and vertical building blocks to implement functionality across or at the same layer.
A modular network stack architecture was studied in \cite{ee2006modular} to avoid code duplication and allow run-time sharing of multiple network protocols.
Their architecture, however, only focuses on the network layer and does not allow for horizontal exchange of building blocks.
%, whereas a generic API across the whole stack is required in the IoT.
The heterogeneous application scenarios of IoT devices require that these building blocks can be arbitrarily combined to a complete network stack.
Very early work on the $x$-Kernel~\cite{hutchinson1988design} provides a common API for protocol composition.

%The most well-known work in this area essentially consists in TinyOS and Contiki network stacks based on IPv6.
%More recent work explored a content-centric network stack for IoT scenarios \cite{icn2014iot} with which, however, IP interoperability is lost.

\paragraph{Loosely Coupled Components}
Many IoT network devices provide their own network stack in addition to software implementations, as well as multiple network stacks may run in parallel on the operating system to dynamically adapt to the deployment environment.
To load protocol functionalities at run time, the network subsystem should loosely combine different building blocks.
All current stacks do not support such feature, either because of data sharing via inflexible callbacks or restricted APIs.

%which allows for merging of IPC and network communication into one API.
%However, as we will show, significant overhead is saved by keeping IPC separate in embedded systems.

%To prevent hardware specific code on the application side, not only a unified API is required but also 

\paragraph{Clear and Consistent APIs}
In particular in the IoT, clean API design is important, as a plethora of technologies exists.
All related network stacks bind names and signatures of functions to technologies.
This makes code technology-dependent, complicates code refactoring, and conflicts with code ergonomics\cite{h-adm-09}.
Our network subsystem implements a key-value-based approach (\path{netapi_dispatch_send(nettype_t, ctx_t, pktsnip_t*)}) that is consistent across all layers.

%
%\begin{verbatim}
%err_t ip6_output(pbuf_t*, ip6_addr_t*, iv6_addr_t*, u8_t, u8_t, u8_t)
%\end{verbatim}

%API design matters~\cite{h-adm-09}.
%APIs cannot be designed without 

%Good APIs are ergonomic.

\paragraph{Comparative Summary of Related Work}
We summarize our observations in Table~\ref{tab:related-work}.
It is worth noting that our comparative analysis of related work is not only based on surveying papers.
Where possible, we also reviewed public code bases to complement open details.
Furthermore, we focus on running the network stacks on constrained IoT hardware, not primarily on border routers.
Consequently, we only indicate support if the corresponding design choice and implementation provides full support of the feature, i.e., horizontal \emph{and} vertical modularity, as well as multi-network interfaces for \emph{more than two devices}.

Our proposed networking subsystem enables our default stack to implement all features.
In the next sections, we show how we achieve this by levering modern system concepts without sacrificing performance on embedded devices.

%even though existing stacks are much more optimized, we show that we can achieve better tradeoffs, richer functionality.

%% file: figs/multi-interface.tex
\begin{tikzpicture}
    \tikzset{
        grass/.style = {decoration={random steps, segment length=2mm}, decorate},
        node/.style = {circle,fill=blue!35,draw=black, minimum width=0.18cm, minimum height=0.18cm, inner sep=0pt},
        gw/.style = {rounded corners=1mm,fill=black!75,draw=black, minimum width=0.7cm, minimum height=0.30cm},
        slant/.style = {yslant=0.5,xslant=-1},
        network/.style = {cloud, cloud puffs=15.7, cloud ignores aspect, minimum width=2cm, minimum height=1cm, align=center, draw},
    }
    \begin{scope}[slant]
        \pgfmathsetseed{1234}
        \draw[black!35,fill=green!20,grass,rounded corners=0.6mm,drop shadow={color=brown!90,shadow yshift=-0.08,shadow xshift=0}] (0,0) rectangle (2.95,2.95);
        \begin{scope}
            \pgfmathsetseed{1234}
            \clip[grass,rounded corners=0.6mm] (0,0) rectangle (2.95,2.95);
            \draw[step=1,dashed,black!35] (0,0) grid (3,3);
        \end{scope}
        \node[inner sep=0pt] (t1) at (2.0,0.5) {\includegraphics[scale=0.30]{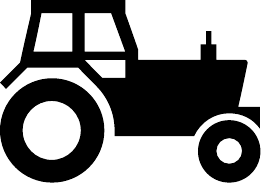}};
        \node[inner sep=0pt] (t2) at (2.5,2.5) {\includegraphics[scale=0.30]{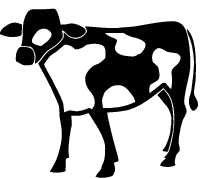}};
        \node[gw] (gw) at (0.25,2) {};
        \draw[black!75,decorate,decoration={expanding waves,angle=30,segment length=0.30cm}] (gw) -- (0.9,0.6);
        \node (r1) at (0.9,1.6) {\includegraphics[scale=0.15]{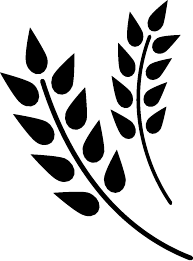}};
        \node (r2) at (0.4,1.1) {\includegraphics[scale=0.15]{figs/plant1}};
        \node (r3) at (0.7,0.6) {\includegraphics[scale=0.15]{figs/plant1}};
        \node (r4) at (0.25,0.5) {\includegraphics[scale=0.15]{figs/plant1}};
        \node (r5) at (0.9,1.1) {\includegraphics[scale=0.15]{figs/plant1}};
     \end{scope}
     \fill[white,draw=black] ([xshift=-1.5mm]gw.center) circle (1.25pt);
     \fill[white,draw=black] ([xshift=-0.0mm]gw.center) circle (1.25pt);
     \fill[white,draw=black] ([xshift=+1.5mm]gw.center) circle (1.25pt);
     \node[network,font=\sffamily] (network) at (-2cm,3cm) {Internet};
     \draw[thick] (gw.west) -| node[font=\sffamily\scriptsize,sloped,above,pos=0.83] {Ethernet} (network.230);
     \draw (gw.north) -- ++(0,0.3) [fill=black] circle (0.4mm) coordinate (gwantenna);
     \draw[slant] (gwantenna) circle (1.4mm) circle(2.5mm);
     \draw (t1.north) -- ++(0,0.3) [fill=black] circle (0.4mm) coordinate (t1antenna);
     \draw[slant] (t1antenna) circle (1.4mm) circle(2.5mm);
     \draw (t2.north) -- ++(0,0.3) [fill=black] circle (0.4mm) coordinate (t2antenna);
     \draw[slant] (t2antenna) circle (1.4mm) circle(2.5mm);
     \draw (gwantenna) to[out=45,in=180] (t2antenna);
     \draw (gwantenna) to[out=35,in=160] (t1antenna);
     \node[font=\sffamily\scriptsize,inner sep=0pt,rotate=-25] at (-0.9,0.10) {IEEE~802.15.4};
     \node[font=\sffamily\scriptsize,inner sep=0pt,rotate=-25] at (0.8,2.8) {LoRa};
     %\node[inner sep=0pt] at ([yshift=-2mm]gw.south west) {\includegraphics[scale=0.025]{figs/riot_logo.png}};
\end{tikzpicture}

%% file: tex/network-stacks.tex
\section{RIOT Networking Subsystem}
\label{sec:nw-stack}

Broad support for networking  is a central building block of RIOT. By design, the system shall be open to flexible layering and varying integration levels of network components. Beyond IP stacks, this is meant to include domain specific technologies, as well as future Internet approaches. Before discussing specific stacks, we first introduce our design rationales and architectural choices.

\subsection{Design Objectives}
\label{sec:netdesign}

The RIOT networking subsystem shall follow a modular architecture with clean interfaces for abstracting all layers.
Our approach is to encapsulate between \one a common hardware abstraction that serves as a unified ground for all network technologies and drivers, and \two a uniform application interface.
Then, network stacks can be easily exchanged or may coexist, when associated with different interfaces.     

In addition, a design goal of our architecture is to provide a common abstraction of protocols that allows for recursive layering and in particular the insertion, exchange, and deletion of any protocol in a particular network stack. Interfaces shall be generic and rich enough to support emerging and future protocols beyond established Internet standards.

These high-level objectives must be implemented with resource frugality and remain compatible with related, highly optimized network stacks to meet the resource constraints of class~1 devices \cite{rfc7228}.
In particular, this has two implications.
First, only small amounts of static memory may be used.
Second, communication tasks between modules and layers must map to the highly efficient IPC of the RIOT kernel in an uncomplex way.
It is worth noting that the RIOT kernel offers a message oriented IPC that allows not only for blocking and non-blocking, but also for synchronous and asynchronous communication.

\subsection{Overall Architecture}
\label{sec:netarch}

\begin{figure}
  \centering
  \includegraphics[width=0.8\columnwidth]{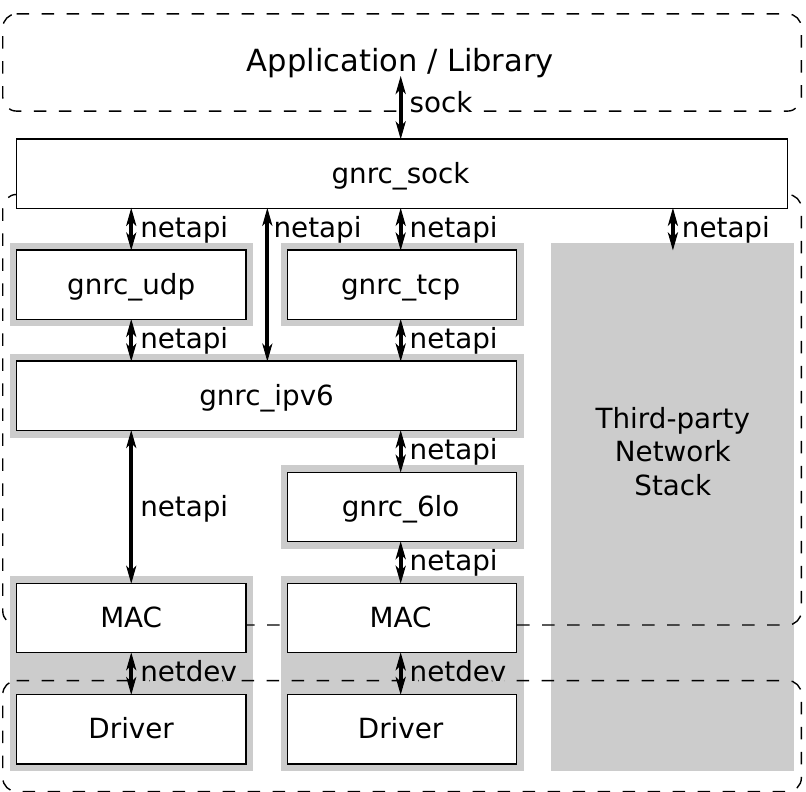}
  \caption{RIOT networking subsystem: recursive architecture with generic APIs.}
  \label{fig:stack_arch}
\end{figure}

The RIOT networking subsystem displays two interfaces to its externals (see Figure~\ref{fig:stack_arch}): The application programming interface \emph{sock} and the device driver API \emph{netdev}. 
Internal to stacks, protocol layers interact via the unified interface \emph{netapi}, thereby defining a recursive layering of a single concept that enables interaction between various building blocks: 6lo with MAC, IP with routing protocols, transport layers with  application protocols, etc. 
This grants enhanced flexibility for network devices that come with stacks integrated at different levels, and provides other advantages as outlined below.
%but in addition we like to emphasize it as an outstanding design feature, 
%In the following, we will outline the core design elements of these APIs.

%\subsubsection{netdev}\label{netdev}
\subsubsection{The Device Driver API: netdev}
 RIOT networking abstracts individual devices via  \emph{netdev} that allows stacks to access network
interfaces in a common, portable way. Unlike common solutions in Linux or other IoT operating systems (e.g., Zephyr, FreeRTOS, Contiki), \emph{netdev} remains fully neutral with respect to the link layer {\em and} the network technology. It exchanges full frames including link layer headers in a buffer provided by the calling stack. The interface does not enforce implementation details concerning memory allocation, data flattening, and threading, but leaves these decisions explicitly to the user of the interface. With only  6 function pointers, \emph{netdev} keeps a very low memory profile. 

The \textit{netdev} interface decomposes into three functional parts: handling of \one network data, \two configuration and initialization, and \three events.
The combination of these three aspects make the interface complete with respect to network device functionality and allow for full control of these devices.

\paragraph{(i) Network data flow}
 \textit{netdev} offers a single {\tt send} function and expects data to be consistent with the link layer in use. This takes the link layer framing out of the device driver context, allowing for more compact device drivers and sharing of code between multiple drivers.

The {\tt send} function is asynchronous and returns as soon as the data has been passed to the network device (e.g., its internal buffer).
For synchronizing on actual transmission the driver can be configured to trigger corresponding {\tt TX\_STARTED} or {\tt TX\_COMPLETE} events as described below.

For most link layer technologies, receiving data is an asynchronous operation.
The \textit{netdev} interface reflects this by splitting the receiving process into two parts, first reacting to RX related events triggered by the network driver, and second reading the actual data from the device.
After a {\tt RX\_COMPLETE} event, the data is available at the network device for the {\tt recv} function. {\tt recv}  offers three different operations: \one getting the size of the incoming data \two reading the incoming packet into a given buffer, dropping the packet afterwards, or \three dropping the incoming data without reading it.
In addition to  incoming network data, the {\tt recv} functions allows to pass a generic data pointer with additional device or link layer specifics such as LQI or RSSI values for wireless devices.

\paragraph{(ii) Configuration and initialization}
The challenge for configuring network devices is the variety of available options, depending not only on the underlying link layer technology, but also on the specific device model and vendor.
\textit{netdev} resolves this by providing generic {\tt get} and {\tt set} functions that read and write all available options, such as link layer addresses, radio channels, or device control  state.
\texttt{netopt}, a global   key-value (meta-)store of options is maintained and extended as needed, each option documenting the type and length of data, as well as the byte order it expects.
It is then up to a device driver to service a given option, or simply return an error code otherwise.

Device initialization  in \textit{netdev} is invoked via the {\tt init} function.
 {\tt init} was chosen to be part of \textit{netdev}, as this gives the network stacks full control over the timing of the initialization, preventing incoming network data or device events to interfer with the setup and initialization of the network stack itself.

\paragraph{(iii) Event handling}
Events triggered by network devices fall into four categories: RX, TX, link, or system related. Correspondingly, four complete sets of event groups are available with \textit{netdev}. 
As not every event fits to every network device, it is up to the device drivers to select the events it supports.
While almost all network device drivers support  {\tt RX\_COMPLTE}  to signal data reception to the upper layer,  only a subset of network devices support particular events such as link layer state changes.
\textit{netdev} foresees the dynamical enabling and disabling of events during runtime, allowing the upper layer to decide which events it is interested in.

Events in RIOT are handled by the interrupt service routine (ISR). In case of an event from a network device, we need to quickly switch out of the ISR to preserve the real-time capabilities of RIOT. For this reason,  \texttt{netdev} allows to register an external event callback which can then switch to a thread context.
Using the \texttt{isr()} function from the thread context, it is able to continue communication with the driver about the event.

%\subsubsection{conn}
%\subsubsection{netapi}\label{netapi}
\subsubsection{The Internal Protocol Interface: netapi}
The RIOT network architecture  defines typed message passing between
 network layers or compound modules with the help of \emph{netapi}.
It was designed to be as simple and versatile as possible, so that even rather  
exotic network protocols can be implemented against it.
Our intention was also to be conveniently modular to make it both easily extensible and
testable.

\emph{netapi} supplies two asynchronous message types \linebreak (\texttt{MSG\_TYPE\_SND}, \texttt{MSG\_TYPE\_RCV}) for packet data communication, and two synchronous message types that expect a reply in form of a \texttt{MSG\_TYPE\_ACK} typed message.
The synchronous message types access \texttt{netopt}, the key-value (meta-)store  already in use by \emph{netdev}.
The reply message must indicate when an option is not supported.
The synchronized message types are:

\begin{itemize}
\item
  \texttt{MSG\_TYPE\_GET} requests an option value from a
  module, while the \texttt{MSG\_TYPE\_ACK} replies  either the length of
  the option value or a negative return to report an error or that the module is not
  supporting the requested option.
\item
  \texttt{MSG\_TYPE\_SET} performs analogous setter operations with a corresponding \texttt{MSG\_TYPE\_ACK} in return.
\end{itemize}

There is no further semantic inherent to messages of the \emph{netapi}, but protocols can require certain
preconditions on packets or option values handed to \emph{netapi} and implement behavior that goes beyond these plain specifications.
This meta-design allows for versatile, yet transparent applications.% of the unified concept.

\begin{table*}
\small
\begin{tabular}{lp{0.81\textwidth}}
\toprule
\textbf{lwIP}~\cite{dunkels2001lwip,fcf2016lwip} &
    Embedded IoT stack that uses an internal IPC API that builds upon centralized message boxes to let both the network device and the application communicate with the central network thread\newline
    \emph{Supporting IPv4, IPv6, 6LoWPAN, TCP, UDP, multiple interfaces}
\\[0.2em]
\textbf{emb6}~\cite{hso2015emb6} &
    Fork of uIP~\cite{dunkels2002uip}, uses a sleep-and-poll scheme on event queues, runs in a single thread\newline
    \emph{Supporting UDP, IPv6, 6LoWPAN, (TCP)}
\\[0.2em]
\textbf{OpenWSN}~\cite{openwsn} &
    Reference implementation of TSCH MAC~\cite{ieee802154, palattella+:2013}, scheduler replaced by RIOT scheduler\newline
    \emph{Supporting 6TiSCH network stack}
\\[0.2em]
\textbf{CCN-lite}~\cite{ccn-lite} &
    Information-centric networking stack with very small core ($\approx$ 1,000~lines of C code). It  uses the GNRC netapi and can be operated on top of any layer that provides a netapi interface\newline
    \emph{Supporting Information-centric networking}
\\[0.2em]
\textbf{NDN-RIOT}~\cite{saz-dinps-16} &
    Optimized port of the Named Data Networking stack. It is built as a single thread on top of netdev and uses the internal IPC API to communicate with applications.\newline
    \emph{Supporting Information-centric networking}\\
\bottomrule
\end{tabular}
\caption{Core characteristics of other network stacks supported in RIOT via the RIOT package and networking system.}
\label{tbl:riot-networkstacks}
\end{table*}

\subsubsection{The User Programming API: sock}
POSIX sockets are the typical south-bound interfaces to communicate with a network stack.
However, POSIX sockets were not made for constrained hardware. They require complex and memory-consuming implementations.
In contrast, \emph{sock} is a collection of high-level network access APIs designed to match the  constraints: \one static memory only, \two highly portable, \three lean and user-friendly.  

Currently supported are \emph{sock\_ip} for raw IP traffic, \emph{sock\_tcp} for TCP traffic, and \emph{sock\_udp} for UDP traffic. Each API can be selected individually by a programmer, thus slimming applications down that only need a part of all. To prevent the need for dynamic memory allocation, the state variable for an end point is always provided via its function call. Portability is ensured by using only common types and definitions either from {\em libc} or POSIX. Applications are thus  easy to port between RIOT and other OSes by wrapping \emph{sock} around the network API of the target OS, e.g., POSIX sockets.

 The GNRC network stack described in the following section implements all three interfaces, 
\textit{sock\_ip}, \textit{sock\_udp}, and \textit{sock\_tcp}.

%% file: tex/gnrc.tex
\section{GNRC and 3rd Party Stacks}\label{sec:gnrc}

As default network stack in RIOT, we introduce the Generic Network Stack (GNRC).
It encapsulates each protocol in its own thread and by following the modular, recursive layering of the RIOT architecture (see Figure \ref{fig:stack_arch}, GNRC makes hot-plugging of a network module possible. 
As such, it resembles UNIX STREAMS~\cite{ritchie1984unix} since it allows chaining of multiple modules by a common API. 
But while STREAMS uses input and output queues to exchange messages between the modules, \emph{GNRC} utilizes the thread-targeted IPC of RIOT with a message queue in each thread. 
The message format is defined by the \emph{netapi}.

\paragraph{netreg} Between modules, the processing chain of a packet is guided by a registry called \emph{netreg}. \emph{netreg} is a simple lookup  table that maps the type ID of a protocol to a list of PIDs. An active thread can look up its succeeding thread(s) using the packet type. 
Modules that are interested in a certain type register with a demultiplexing context like a port in UDP. If a thread wants to send a message to a certain UDP port, it looks up the
protocol type  in the \emph{netreg} and then searches for the demultiplexing context, e.g., the requested port in UDP.
Every PID in the registry that meets both of these parameters, will  be a target
for the message, e.g., containing the UDP packet with this port.

\paragraph{Packet Buffer} GNRC uses a flexible central packet buffer \emph{pktbuf} to store packet components during stack traversal. \emph{pktbuf} handles chunks of variable length and uses this to operate on fractions of packets that are called ``snips''. Snips are usually used to differentiate between different headers and the payload. They can be re-used within the buffer, thus allowing for de-duplication of (mainly) header space. This flexible storage  saves GNRC from internal~copying~of~data, provides efficient and structured access to protocol infor\-mation, and is of particular convenience for complex protocol functions like fragmentation and retransmission. It is noteworthy that the short lifetime of packets in the buf\-fer and its even distribution of header sizes makes external fragmentation unlikely, which we empirically confirmed.

For thread-safety, a packet snip keeps track of  threads using it via a
reference counter. If a thread
wants to write to the packet buffer and the reference counter is larger than
one, the packet snip and all its descendants are duplicated in a 
\emph{copy-on-write} style. To keep duplication minimal,
only snips reachable from the current head are copied, assigning a tree structure to a packet
 in the packet buffer. For this reason, the order of  packet snips  in reception is
reversed---starting with the payload, ending with the header of the lowermost
layer---while a packet in transmission is stored in forward order.

 The advantages of storing packet snips in this central buffer are 

\begin{itemize}
    \item \emph{saving memory} by keeping the length of the packet 
        dynamically allocated instead of using fixed-sized chunks that might not
        be used to their full size
    \item \emph{avoiding duplication along the stack}, since every protocol implementation
        (including TCP) is able to use the snips instead of keeping its own buffer.
    \item \emph{handling multiple packets} as one, as the number of
        packets is not directly linked to the size of the buffer.
\end{itemize}

\paragraph{Other Stacks} \label{sec:onw-stacks} 
Several third-party network stacks have been ported and integrated into the RIOT architecture. Some are well-known IP stacks from Wireless Sensor Nets, some are early developments of future networking technologies. We give a comprehensive overview in Table~\ref{tbl:riot-networkstacks}.

%% file: tex/evaluation.tex
\section{Evaluation}
\label{sec:neteval}

We now evaluate the performance of RIOT networking and quantify the cost of each component in detail. 
The modular RIOT architecture in general, but in particular its cleanly layered, multithreaded generic network stack GNRC introduce some overhead. We want to answer  questions about its general feasibility and overall communication performance. We will also contrast the efforts taken at each individual layer with its merits, and evaluate the price to pay for user friendly POSIX interfaces.  
%To quantify this all, we will dive deeply into the networking software and measure all relevant elements in specific microbenchmarks. We examine the metrics {\em throughput, goodput}, {\em packet processing rates},  {\em processing times per packet}, as well as the  {\em consumptions of memory and energy}. Unless noted otherwise, all measurements were performed on the iotlab-m3 hardware as described in the previous section. 
To quantify all this in detail, we conduct micro-brenchmarks and examine the metrics {\em throughput, goodput}, {\em packet processing rates},  {\em processing times per packet}, as well as the  {\em consumptions of memory and energy}. Unless noted otherwise, all measurements were performed on the iotlab-m3 node as described in the previous section. 

After evaluating the GNRC stack performance in detail, we draw comparisons to other stacks from the IP family, as well as from Information-centric Networking (ICN). By integrating popular network stacks into RIOT, we demonstrate \one the flexibility of our networking architecture, and \two create identical environments for the different network softwares. This exhaustive performance examination is completed by comparative host-to-host measurements between RIOT and Contiki.

%\subsubsection{Network Software}
\subsection{Benchmarking Software Layers}

\begin{figure}
  \centering
  \includegraphics[width=1.0\columnwidth]{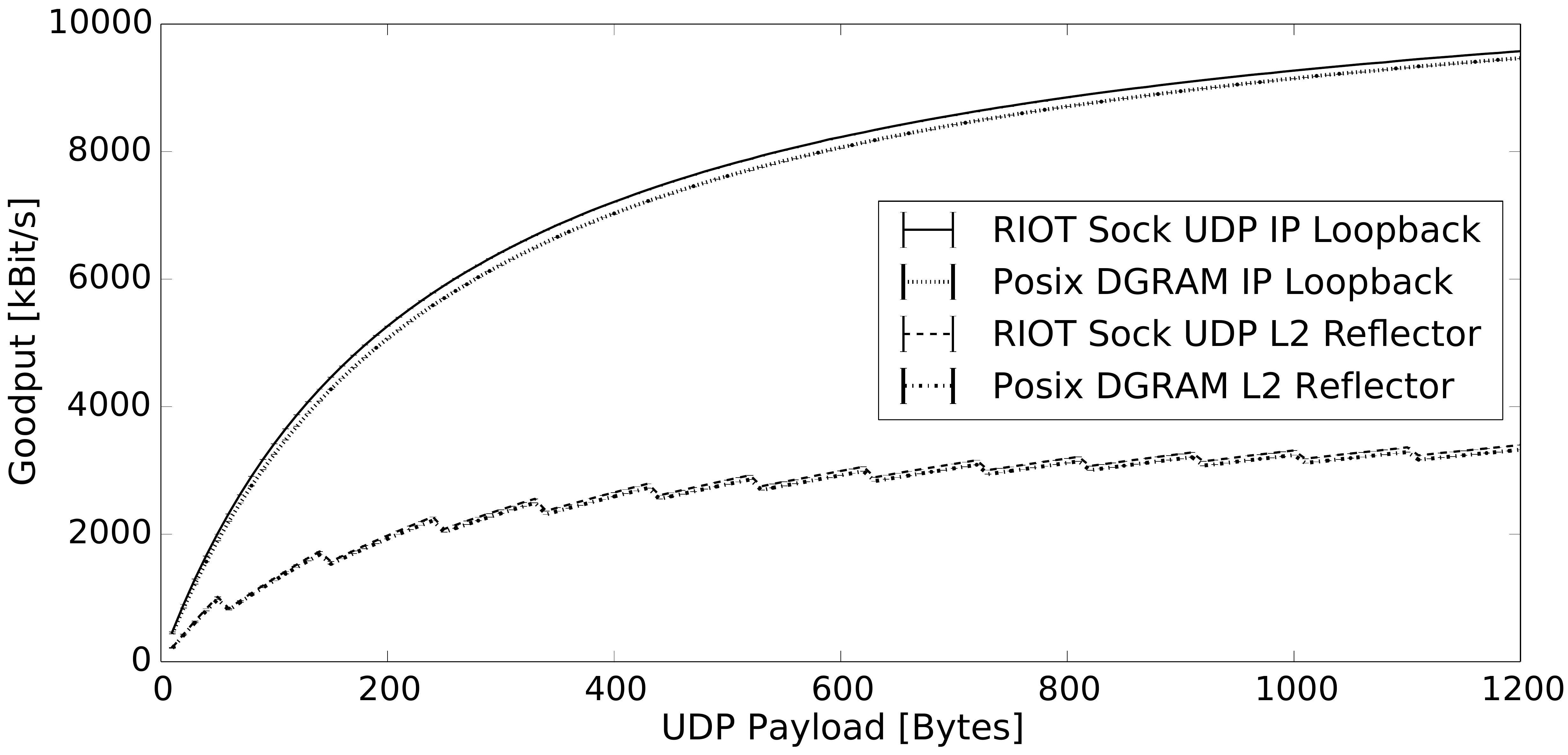}
  \caption{Effective stack throughput in a loop---bidirectional traversal w/ and w/o 6Lo and POSIX.}
  \label{fig:throughput_stack_udp}
\end{figure}

\begin{figure*}
  \centering
  %\subfigure[Average processing times per packet---cumulative CPU overhead of the different network layers.]{\includegraphics[width=1.0\columnwidth]{processing_times_posix_all}
  \subfigure[Average processing times per packet.]{\includegraphics[width=1.0\columnwidth]{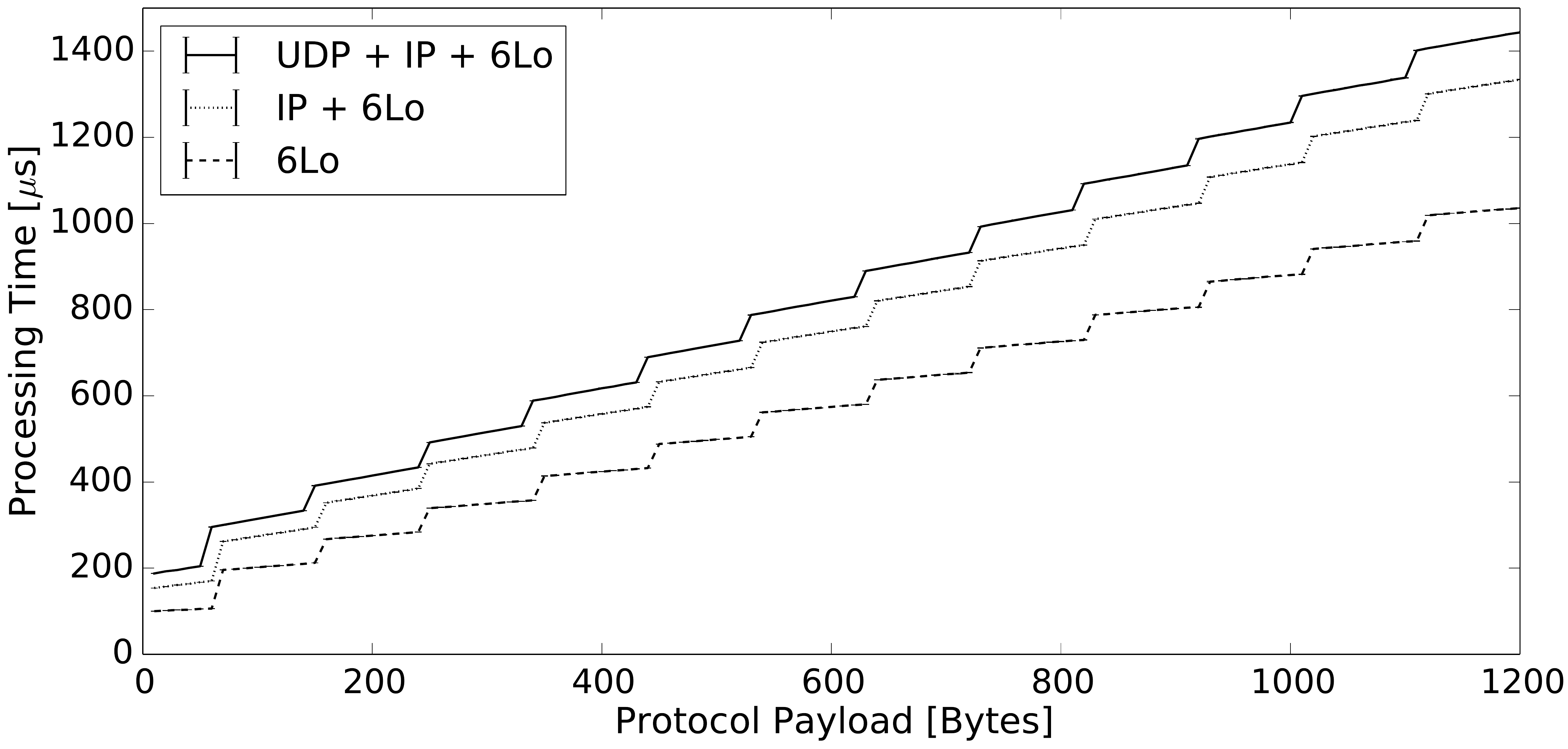}
  \label{fig:processing_stack_all}
  }
  \subfigure[Packet processing rates.]{\includegraphics[width=1.0\columnwidth]{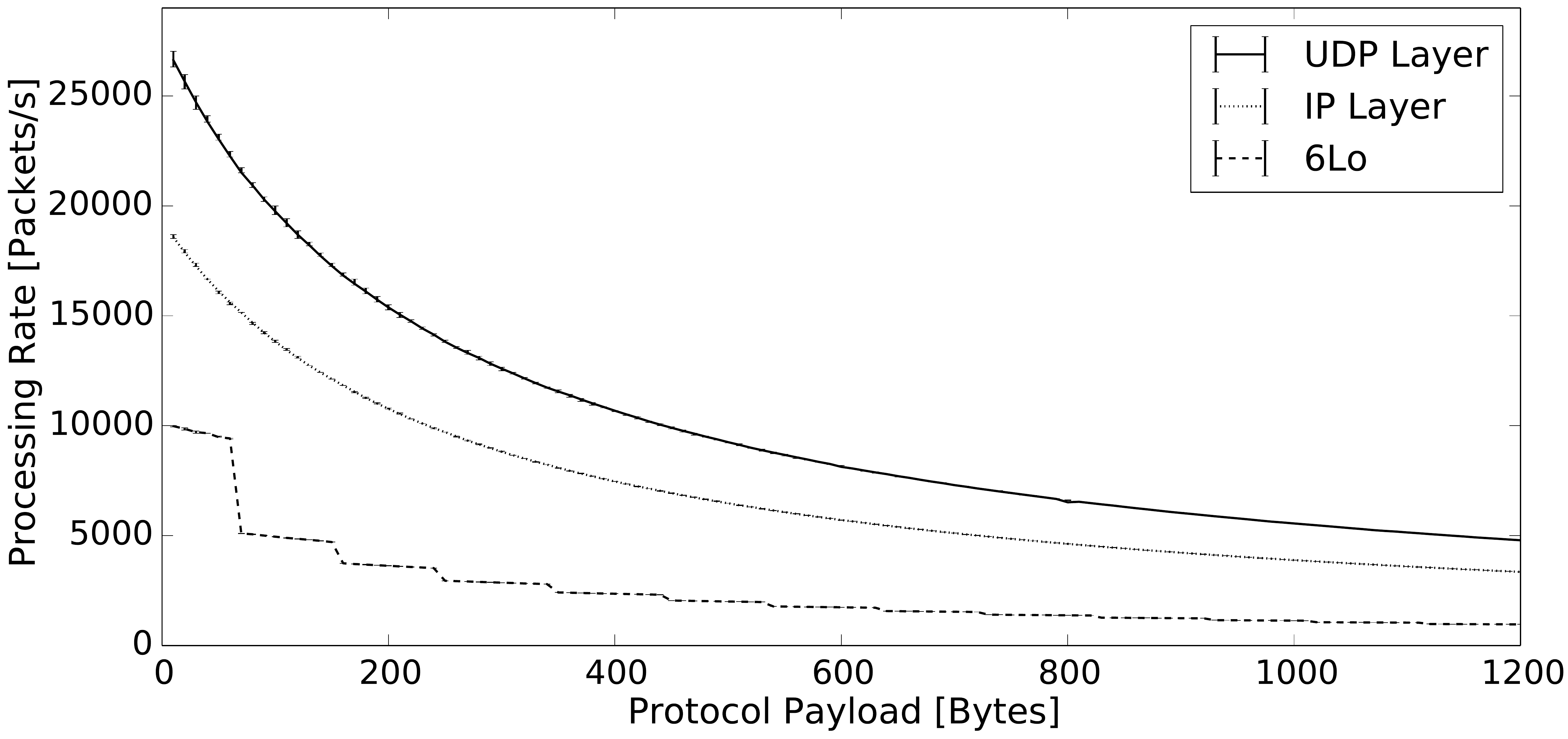}
  \label{fig:rate_stack_all}}
  \caption{Cumulative CPU overhead of the different network layers and processing capacities per network layer.}
  \label{fig:processing}
\end{figure*}

Our first goal is to empirically measure the different layers of the RIOT networking software. For this, we employ a simple network testing program that transmits and receives packets down and up the network stack, using \one the IP loopback and \two a minimal packet reflector that serves as loopback on layer~2. Note that this evaluation does not include device drivers. By issuing packets at the different end points, i.e., the different APIs for UDP and raw IP packet generation, we are able to quantify efforts of the stack traversal for all dedicated layers of the network stack.

\paragraph{Goodput}
The effective data throughput of an UDP application is displayed in Figure~\ref{fig:throughput_stack_udp}. Results show  clearly divergent performance for the IP loopback and the L2 virtual reflector device, which is mainly the result of the 6LoWPAN fragmentation. Fragmentation is clearly visible in the step function of the reflector curve. This means that---independent of the physical network in\-ter\-face---the RIOT network stack is able to serve (unidirectional) communication of 10 Mbit/s on the IPv6 layer, whereas 6LoWPAN packet fragmentation reduces the throughput to about a few Mbit/s.  Correspondingly, a low-end RIOT node is well able to participate in an 802.11a/b base network at full speed, while it also proves feasible to serve a LoWPAN of tiny MTU up to transmissions in the range of Mbit/s. Comparisons of the different user APIs show little effect ($< 3 \%$). 

\paragraph{Processing Overhead per Software Layer}
Processing efforts of the network software are further differentiated in Figure \ref{fig:processing_stack_all}. Splitting a 1,000~Bytes packet into 11~fragments or reassembling it requires about 800~$\mu s$ by the 6Lo layer, while IPv6 and UDP require only about  200~$\mu s$ and  50~$\mu s$ respectively. It is noteworthy that the complexity of IP strongly depends on the direction of software layers. For outgoing packets, IP has to perform next hop determination (including FIB lookup and neighbor cache management), while incoming packets are simply demultiplexed towards upper layers. Our measurements of symmetric stack traversal do not distinguish, but average over directions. The remaining efforts at the UDP layer basically concern checksum calculations and verifications respectively.

These results lead to largely varying packet processing rates of the different software layers (see Figure~\ref{fig:rate_stack_all}). While UDP can handle almost 10,000 packets of 1,000~Bytes per second, 6Lo operates one order of magnitude slower. Nevertheless, all results are clearly in line with the underlying protocol complexity, and the smooth measurement curves reflect the absence of disturbing implementation artifacts.

%\begin{figure}
  %\centering
  %\includegraphics[width=1.0\columnwidth]{processing_rate_all}
  %\caption{Packet processing rates at the  software layers.}
  %\label{fig:rate_stack_all}
%\end{figure}

\begin{figure}
  \centering
  \includegraphics[width=1.0\columnwidth]{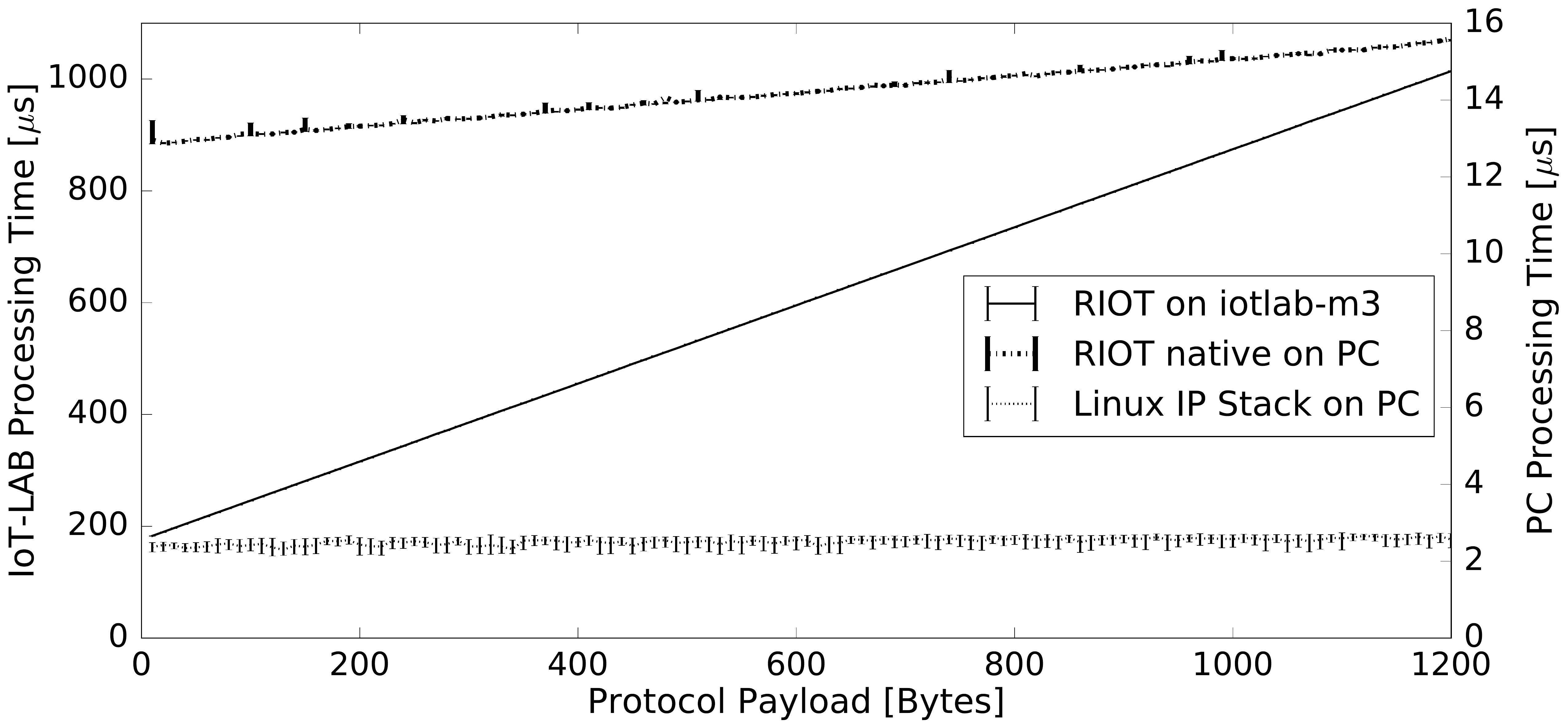}
  %\caption{Scaling the stack---packet processing times on a Linux PC.}
  \caption{Comparing packet processing times between RIOT on real hardware (left ordinate), emulated on Linux and Linux (right ordinate).}
  \label{fig:stackspeed_riot_linux}
\end{figure}

\paragraph{Upscaling: RIOT versus Linux}
Judging on network software, a relevant question arises about the scalability of the stack.
Since the performance discussed so far was mainly limited by the low-end iotlab-m3 system, we extend our measurements to a current desktop PC (Intel i5-3570, 3.4 GHz), deploying RIOT native.
RIOT \emph{native} runs the same RIOT code base virtualized as a Linux process on this machine, which operates at a 50 times faster clock speed.
Due to large memory caches, enhanced buses etc. on the PC, we expect software to perform more than 50 times faster on this host.

Results for packet processing times of the RIOT IoT node, RIOT \emph{native}, and of the native Linux IP stack are compared in Figure \ref{fig:stackspeed_riot_linux}. Measurements for RIOT \emph{native} show a base overhead in the order of 10~$\mu s$ which can be attributed to the emulation environment. Network processing times grow on a scale of 3~$\mu s$ (13--16~$\mu s$)---in contrast to  700~$\mu s$ on the iotlab-m3 board. As expected, these values indicate a relative speed-up in the order of 100, when emulation overhead is disregarded. Comparing RIOT \emph{native} emulation with the native Linux UDP/IP stack, the overall performance gap is a factor of six. While Linux can process more than 500,000 packets per second,  RIOT \emph{native} almost reaches 100,000---we consider this excellently scalable.

%% TODO for final: Quantify emulation overhead

\paragraph{Memory}
Next we evaluate memory consumption of the GNRC stack components. Table~\ref{tab:stack_memtest} displays usages of ROM and RAM for our test applications that include selected parts of the network stack. We compare the allocated resources of a simple UDP communicator using the network stack (including its universal packet buffer) to speak down to the Layer 3, down to 6Lo, and via the 802.15.4 device driver. 

In these lean, real-world experiments we find values from the generic range derived in Section~\ref{sec:background}. This shows that the RIOT  network architecture with its re-usable buffer and recursively layered stack closely complies to real-world applications. Additionally, we quantify the overhead of the POSIX API to be rather small ($ \sim 3 \%$). 

\begin{table}
\begin{center}
\begin{tabular}{l c c}
\toprule
\textbf{Application} & \textbf{ROM} & \textbf{RAM} \\
\midrule
Sock UDP -- IP Loopback & 26,444 & 13,192 \\
Sock UDP -- L2 Reflector & 35,880 & 15,112 \\
Sock UDP -- IF Driver & 38,912 & 15,151 \\
\midrule
POSIX API Overhead & 1,260 & 500 \\
\bottomrule
\end{tabular}
\caption{Memory consumptions [Bytes] for the test applications for varying stack depths (w/o application buffers).}
\label{tab:stack_memtest}
\end{center}
\end{table}

\iffalse

All RIOT system components pre-allocate memory statically, which raises the question about actual RAM utilisation. We analyzed the real memory consumption within the threaded layers (excluding basic structures like the packet buffer).  For this, we flagged memory prior to consumption and counted untouched addresses after processing. Relative results are shown in Table \ref{tbl:mem_stack_all} and indicate that two thirds of the allocated memory remain unused. The reasons behind are generic thread allocations at each layer, which include a fixed context data structures (1,024 Bytes). Future revisions of the GNRC stack might tailor thread parametrization and thereby activate this memory-cutting potential.

\begin{table}
\small
\begin{tabular}{p{0.4cm}p{0.5cm}p{1.0cm}p{0.45cm}p{0.45cm}p{0.450cm}p{0.4cm}p{0.75cm}}
\toprule
\textbf{Idle} & \textbf{Main} & \textbf{UDP~Srv} & \textbf{UDP} & \textbf{IPv6} & \textbf{6Lo} & \textbf{~~L2} & \textbf{Unused} \\
\midrule
1.9\% & ~~~9\% & ~~5.9\% & 5.6\% & 4.6\% & 5.5\% & 5.4\% & 62.1\% \\
\bottomrule
\end{tabular}
\caption{Relative distribution of memory consumption within the allocated range of the test application.}
\label{tbl:mem_stack_all}
\end{table}
\fi

\begin{figure}
  \centering
  \includegraphics[width=1.0\columnwidth]{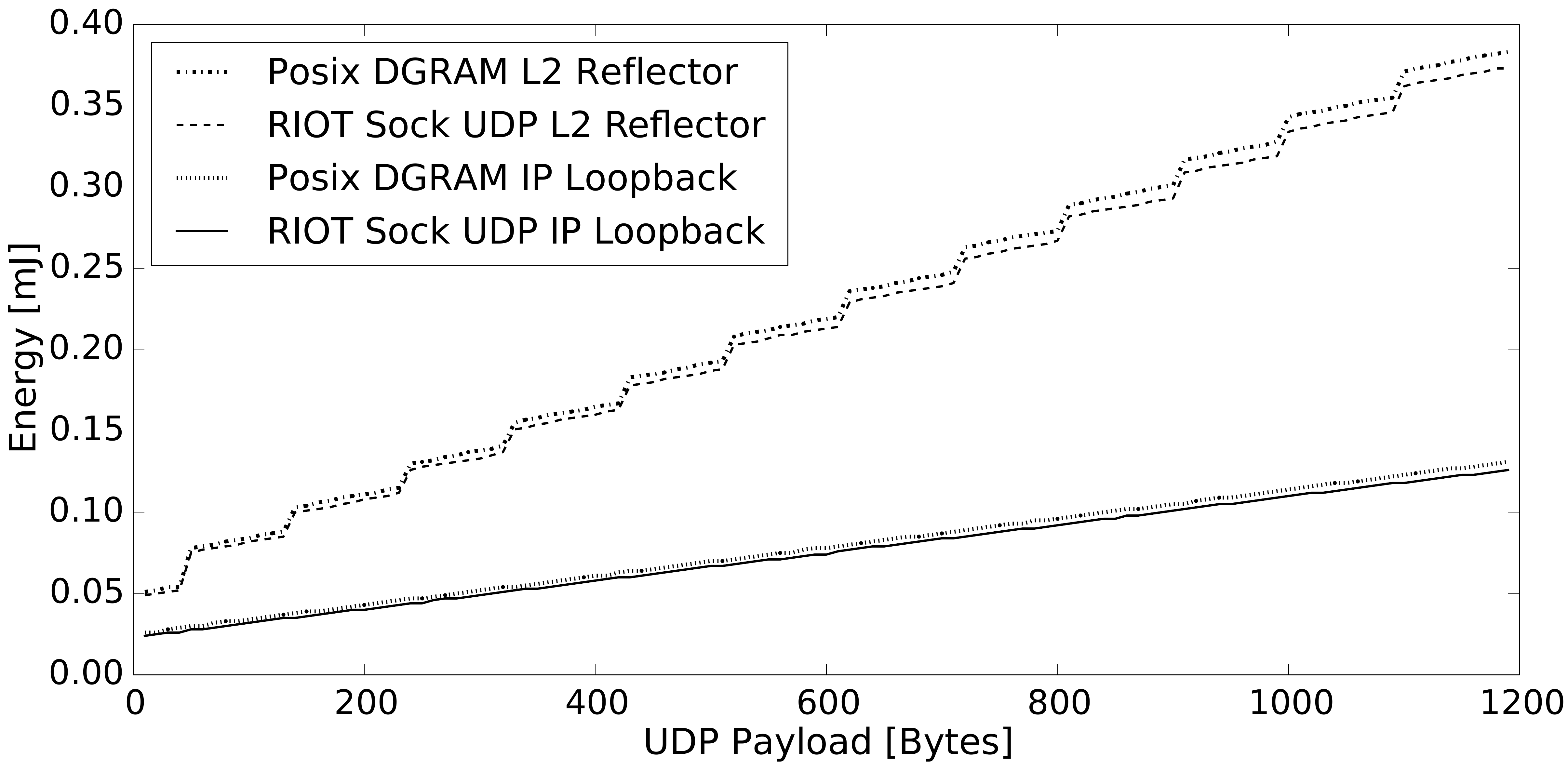}
  \caption{Energy consumption per UDP packet w/ and w/o 6Lo and POSIX}
  \label{fig:energy_stack_all}
\end{figure}

\paragraph{Energy}
Our final microbenchmark of the software stack examines energy consumption. The iotlab-m3 board runs at about 90~$mW$ without staged power saving modes. Hence our results in Figure \ref{fig:energy_stack_all} directly correspond to processing times. 

\subsection{Measuring the Full Stack with Radio}

Next we want to explore the full stack  in wireless operation, which includes the Atmel AT86RF231 transceiver with its driver. We move to an EMC (electromagnetic compatibility) cleanroom to prevent unwanted interferences from neighboring radio sources.
We transfer UDP packets at maximum speed from and to the RIOT test system by using a Raspberry PI peer (with AT86RF233 Transceiver)---this  counterpart serves the purpose of fully adapting its predominant  resources to the IoT board while enforcing network compatibility at the same time.  

\begin{figure*}
  \centering
  \subfigure[Transmission]{\includegraphics[width=1.0\columnwidth]{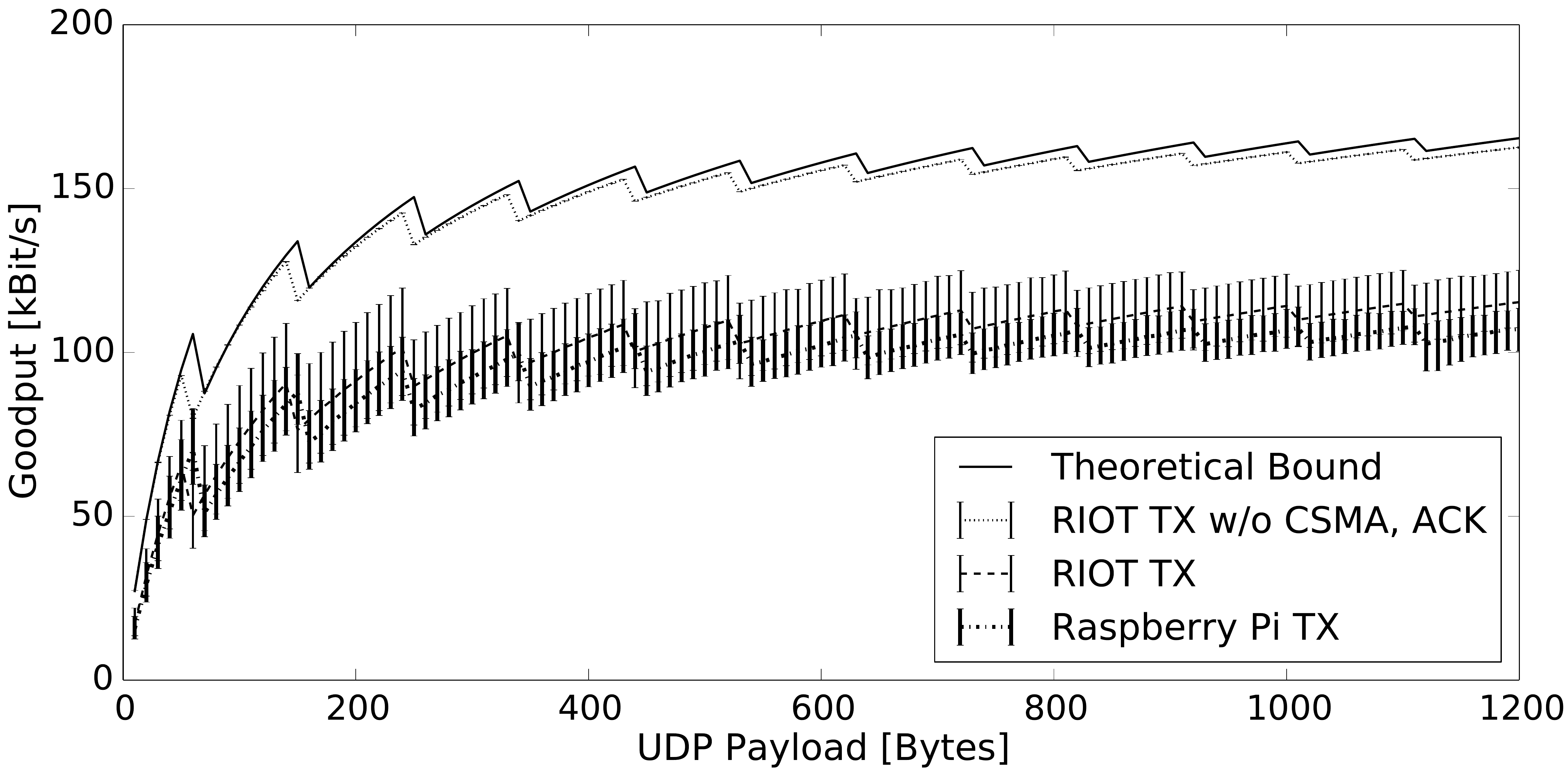}
  %\caption{Effective data transmission with 802.15.4 transceiver versus theory}
    \label{fig:throughput_transmission_tx}
  }
  \subfigure[Reception]{\includegraphics[width=1.0\columnwidth]{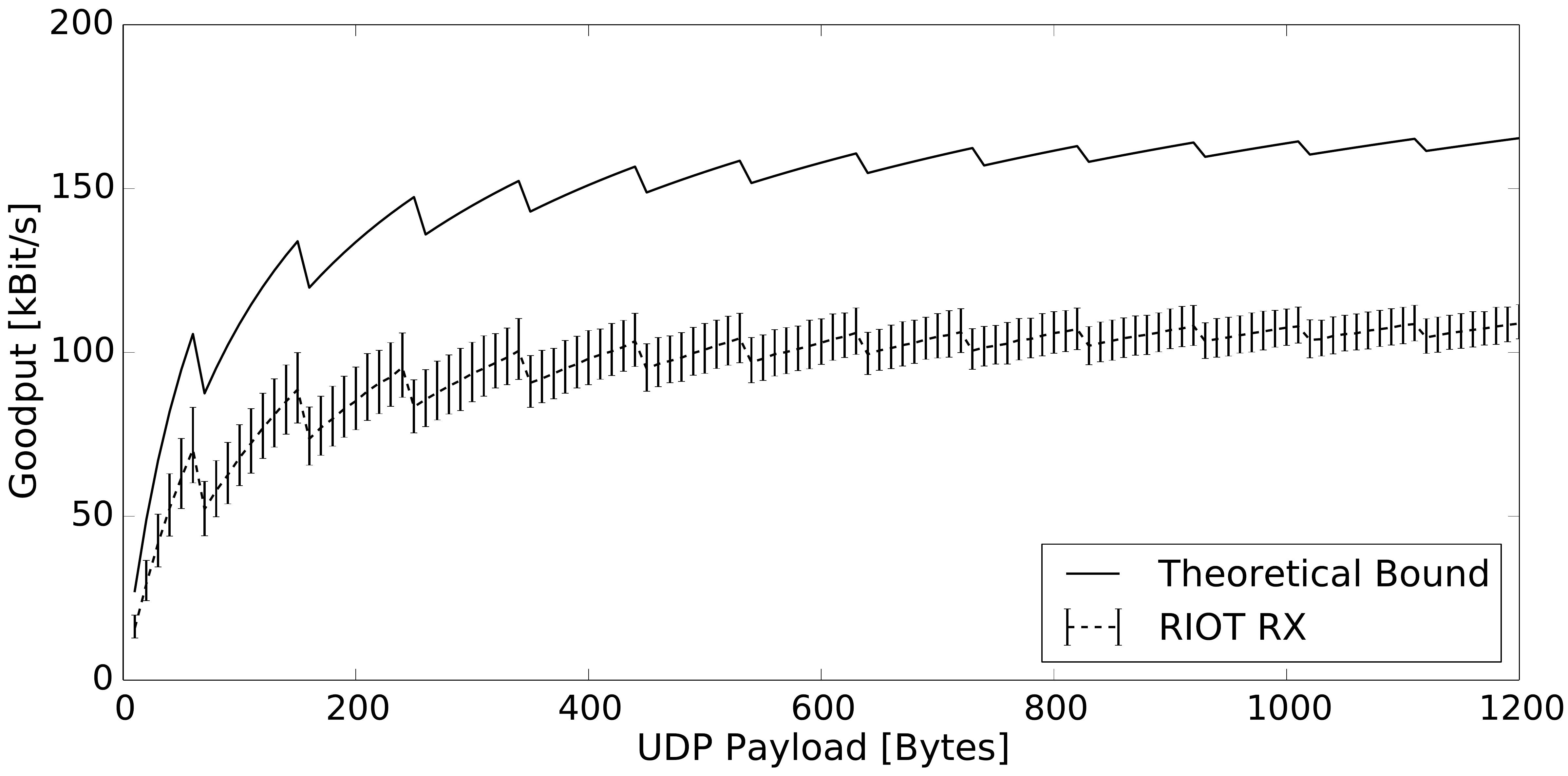}
  %\caption{Effective data reception with 802.15.4 transceiver versus theory}
  \label{fig:throughput_transmission_rx}
  }
  \caption{Effective goodput with 802.15.4 transceiver versus theory}
  \label{fig:throughput_transmission}
  %\vspace{-0.5cm}
\end{figure*}

\paragraph{Sender Goodput}
Figure~\ref{fig:throughput_transmission_tx} compares the UDP goodput of the packet transmission at the RIOT node with its theoretic upper bound. The latter was calculated based on the plain transfer capacity of the PHY, accounting for headers and the long and short interframe spacing (LIFS=40 and SIFS=12 symbols), but disregarding all contention delays. 

Results remain below two third of the theoretical capacity, which however is due to the MAC layer coordination by CSMA and acknowledgements of frames. Actual throughput nicely approaches its limits after these MAC layer functions have been disabled (see the dotted line in Figure~\ref{fig:throughput_transmission_tx}). These results indicate that the RIOT stack serves the interface driver at full speed. This is consistent with measurements of the plain software stack (see Figure~\ref{fig:throughput_stack_udp}) which yield a  processing throughput  about 50 times higher than 802.15.4 speed. To confirm, we also compare with the transmission of the PI---finding even slightly lower throughput. 

\paragraph{Receiver Goodput}
The performance of the RIOT receiver accurately mirrors the transmission capacities as displayed in Figure~\ref{fig:throughput_transmission_rx}. These results clearly reflect that the overall network throughput is determined by the 802.15.4 network access, while the RIOT network software does not slow down communication anywhere. It is worth noting that receiver measurements admit exceptionally low error bars, which is due to the clean EMC environment. Side channel interferences may be large in the wild, and real-world throughput varies on a more pronounced scale.

%\begin{figure}
  %\centering
  %\includegraphics[width=1.0\columnwidth]{CABINE_throughput_transmission_rx}
  %\caption{Effective data reception with 802.15.4 transceiver versus theory}
  %\label{fig:throughput_transmission_rx}
%\end{figure}

\begin{figure}
  \centering
  \includegraphics[width=1.0\columnwidth]{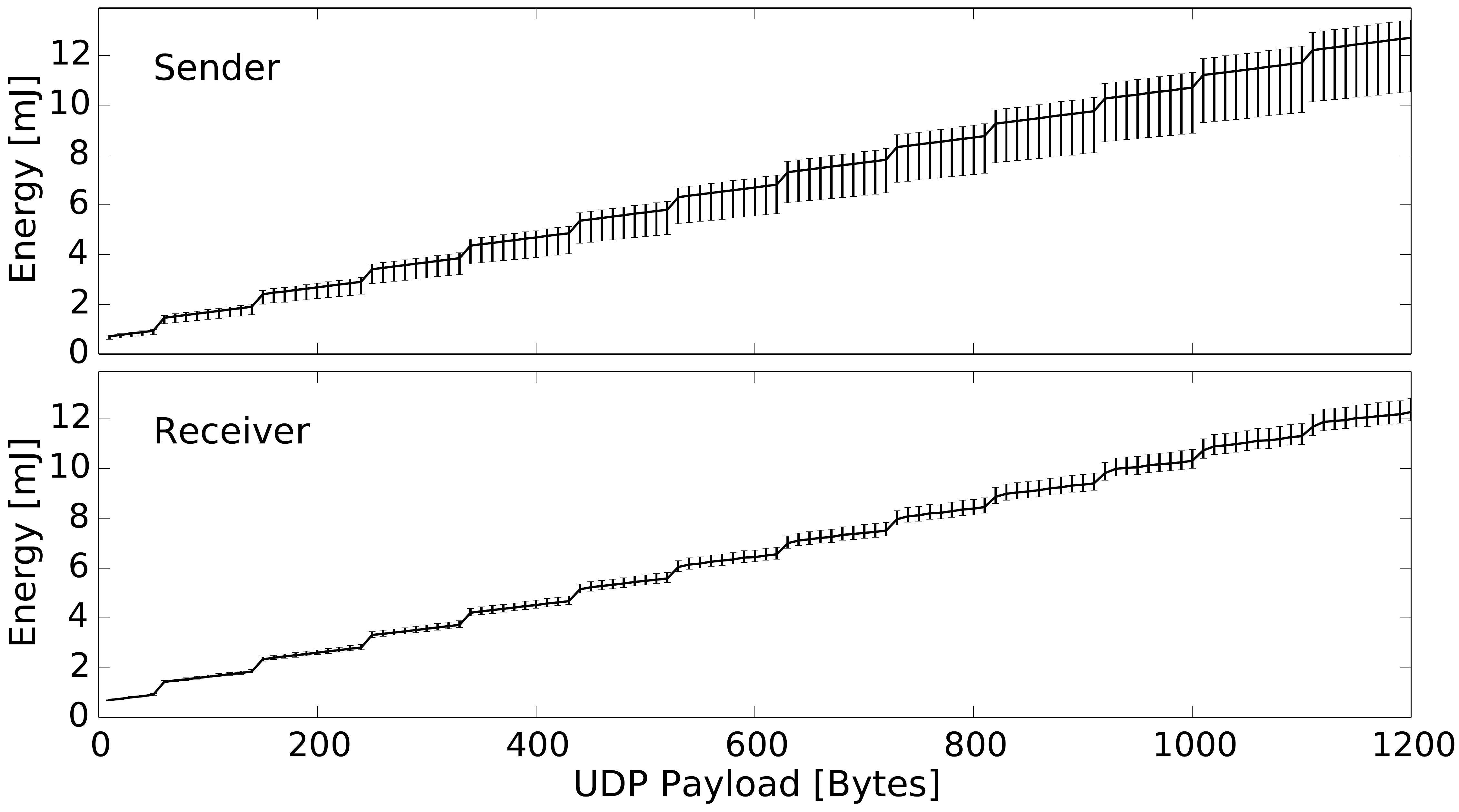}
  \caption{Energy consumption per UDP packet during wireless transmission at sender and receiver}
  \label{fig:energy_transmission_rx-tx}
\end{figure}

\paragraph{Energy}
The overall energy consumption which includes packet processing and radio transmission/reception is displayed in Figure~\ref{fig:energy_transmission_rx-tx}. Results exceed measurements of the pure stack by a factor of 30---plainly due to wireless transmission.  With $\approx 45\, mW$, the Atmel transceiver accounts for about half the power of the system, and wireless transmission was seen $\approx 50$ times slower than the pure stack.  

%% TODO for final: Quantify energy just consumed by the transceiver (w/o driver overhead)

\subsection{Comparing GNRC with Other Stacks}\label{sec:eval-compare}

\subsubsection{Embedded IP Stacks}

In the following performance evaluation of our stack, we compare GNRC with the other IoT IP stacks introduced in Section \ref{sec:onw-stacks} that have been ported to RIOT, namely lwIP, emb6 (w/o RPL). Comparisons are drawn to the corresponding GNRC version (w/o RPL). Buffer configurations of all stacks were chosen to fit one full IPv6 packet including 6LoWPAN fragmentation meta data. All measurements continue to be performed on the iotlab-m3 hardware. 

\begin{figure}
  \centering
  \subfigure[ROM]{\includegraphics[width=0.48\columnwidth]{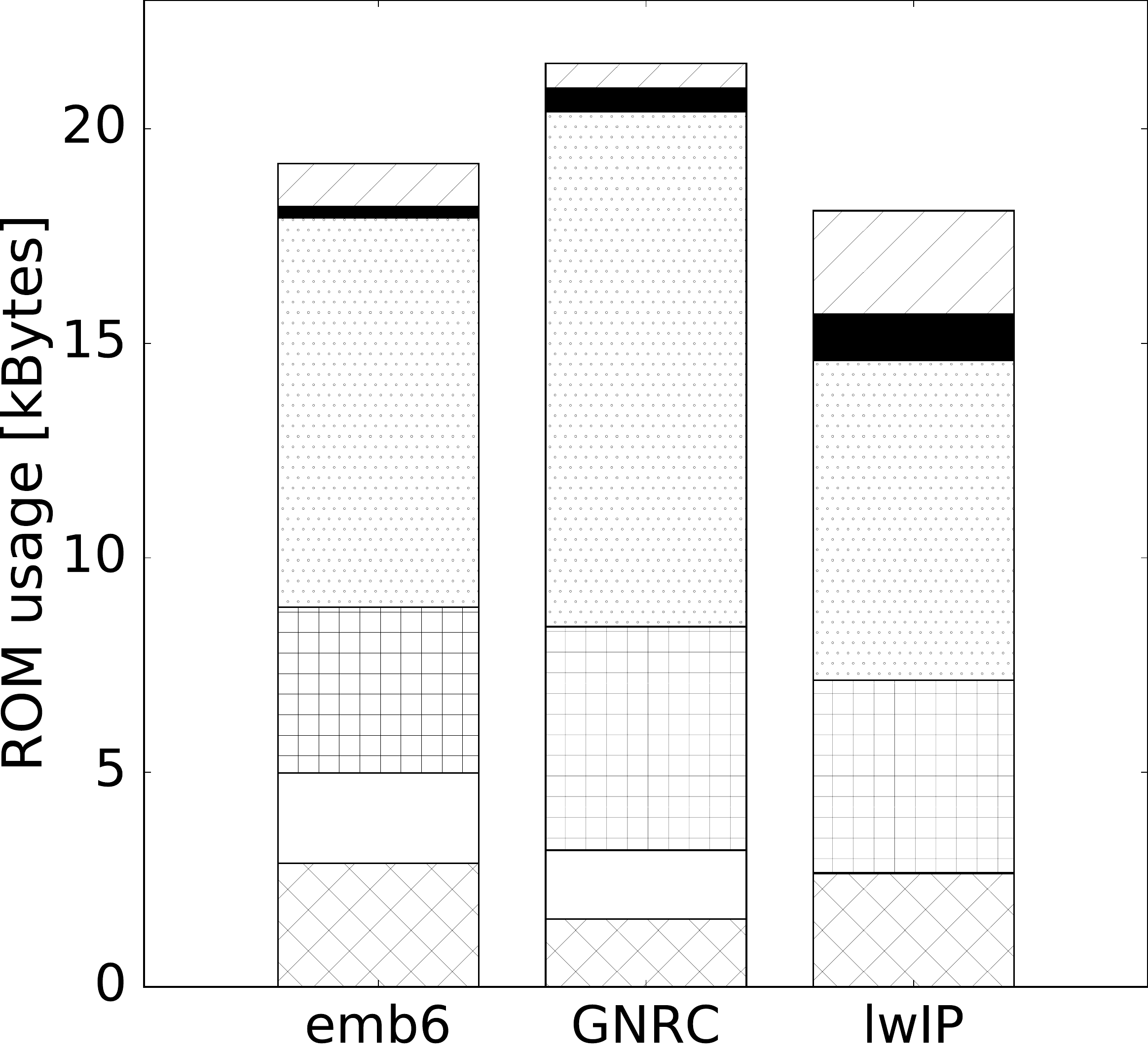}
  %\caption{Effective data transmission with 802.15.4 transceiver versus theory}
   % \label{fig:nw_comp_ram}
  }
  \subfigure[RAM]{\includegraphics[width=0.48\columnwidth]{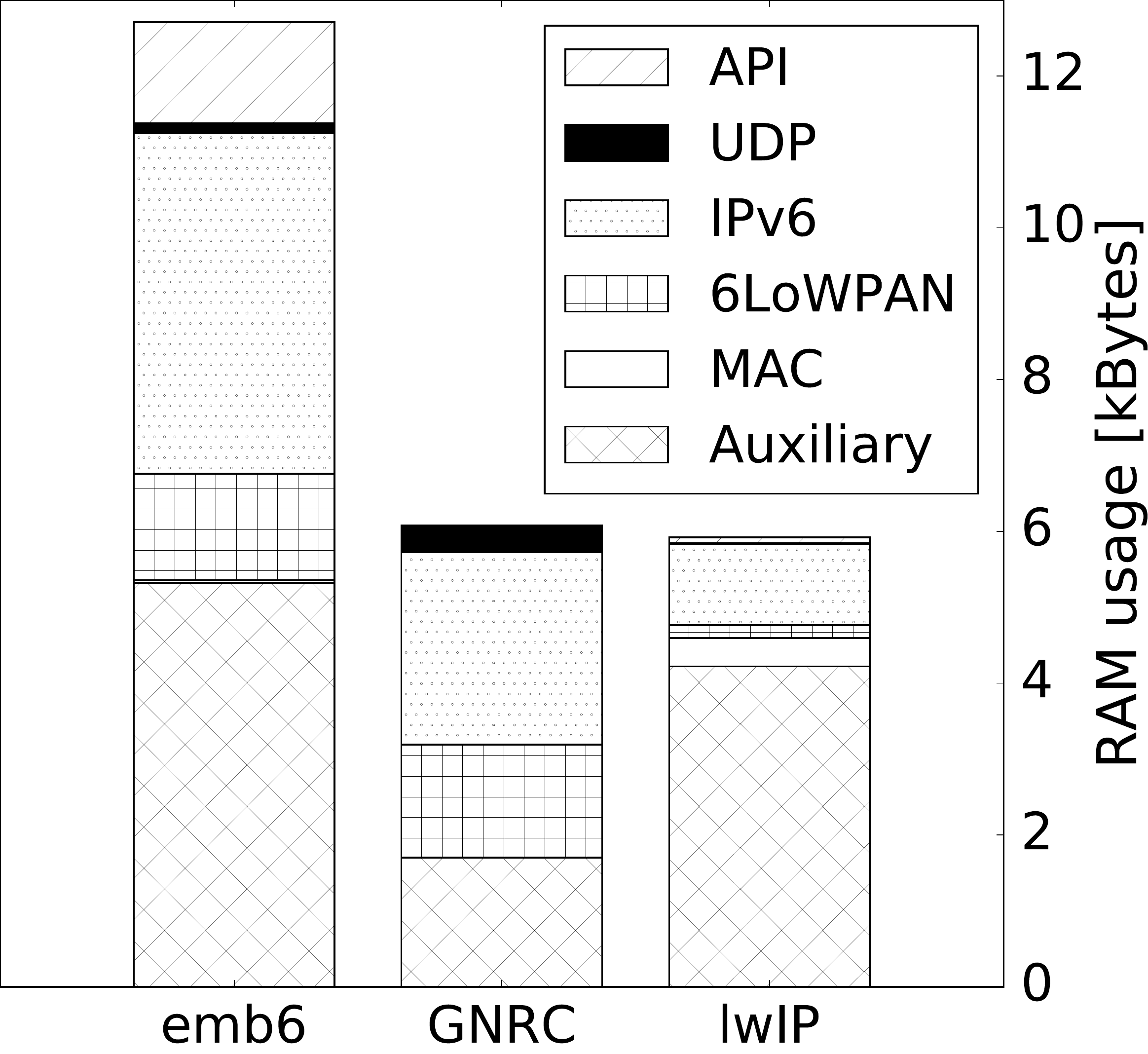}
  %\caption{Effective data reception with 802.15.4 transceiver versus theory}
  %\label{fig:nw_comp_ram}
  }
  \caption{Sizes of different IP network stacks on RIOT}
  \label{fig:nw_comp_rom_ram}
%  \vspace{-0.28cm}
\end{figure}

\paragraph{Memory} We compare ROM and RAM sizes of the stacks in Figure \ref{fig:nw_comp_rom_ram}. Clearly GNRC admits the largest code size, exceeding emb6 and lwIP by 3--4 kB only, though. Given that the mature implementations of lwIP and emb6 meet the very young code of GNRC, this comes at little surprise. Code optimisations can be expected in the future, in particular to decrease the sizes of IPv6 and 6LoWPAN which are dominant in all implementations.

RAM consumption of GNRC remains low ($\sim 6 kB$) slightly exceeding  lwIP by fewer than 100 Bytes. GNRC takes advantage of its central, universal packet buffer for variably-sized chunks. All other stacks consume significantly larger auxiliary data storage, lwIP by its use of fixed size chunks tailored to maximal needs. It is noteworthy that the GNRC measurements subsume the full API abstractions netdev and netapi which are not part of the competing stacks.

\begin{figure}
  \centering
  \includegraphics[width=1.0\columnwidth]{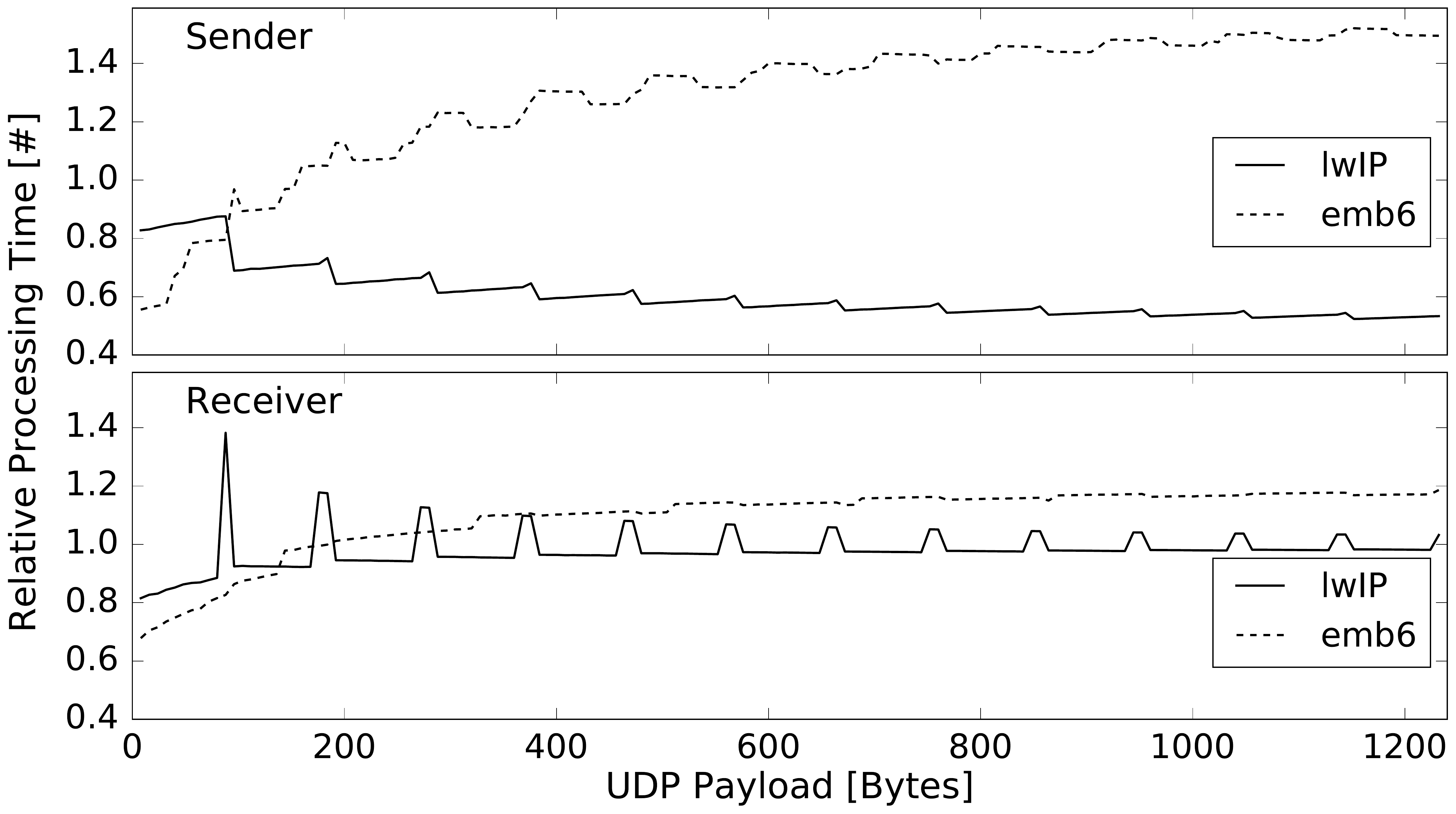}
  \caption{Runtime of networks stacks relative to GNRC}
  \label{fig:nw_comp_proc_time}
\end{figure}

\paragraph{Processing Times} The relative runtime performance of the stacks with respect to its matching GNRC version is depicted in Figure~\ref{fig:nw_comp_proc_time}. Traversal times for the emb6 stacks are significantly slower than GNRC ($\sim 20 - 40\, \%$) except for tinygrams. This is likely caused by slow packet queues for fragment processing. The receiver performance of lwIP is comparable to GNRC. Peaks appear because fragment processing is not done for optimal payload sizes---a known issue in lwIP. IwIP uses an {\tt mbox-}based IPC messaging that is currently much slower than the thread-targeted IPC in RIOT, but GNRC on the other hand uses multiple IPC calls between its layers. It is still faster than the purely function-, but also sleep-cycle based emb6.

However, lwIP outperforms GNRC for packet transmission by up to 40 \%. This superior sender performance is the result of an effective thread blocking by the sending task. lwIP fragments and sends a datagram in one go without yielding the processor, thereby paying the fast transmission with a complete block of the reception.   

\begin{figure}
  \centering
  \includegraphics[width=1.0\columnwidth]{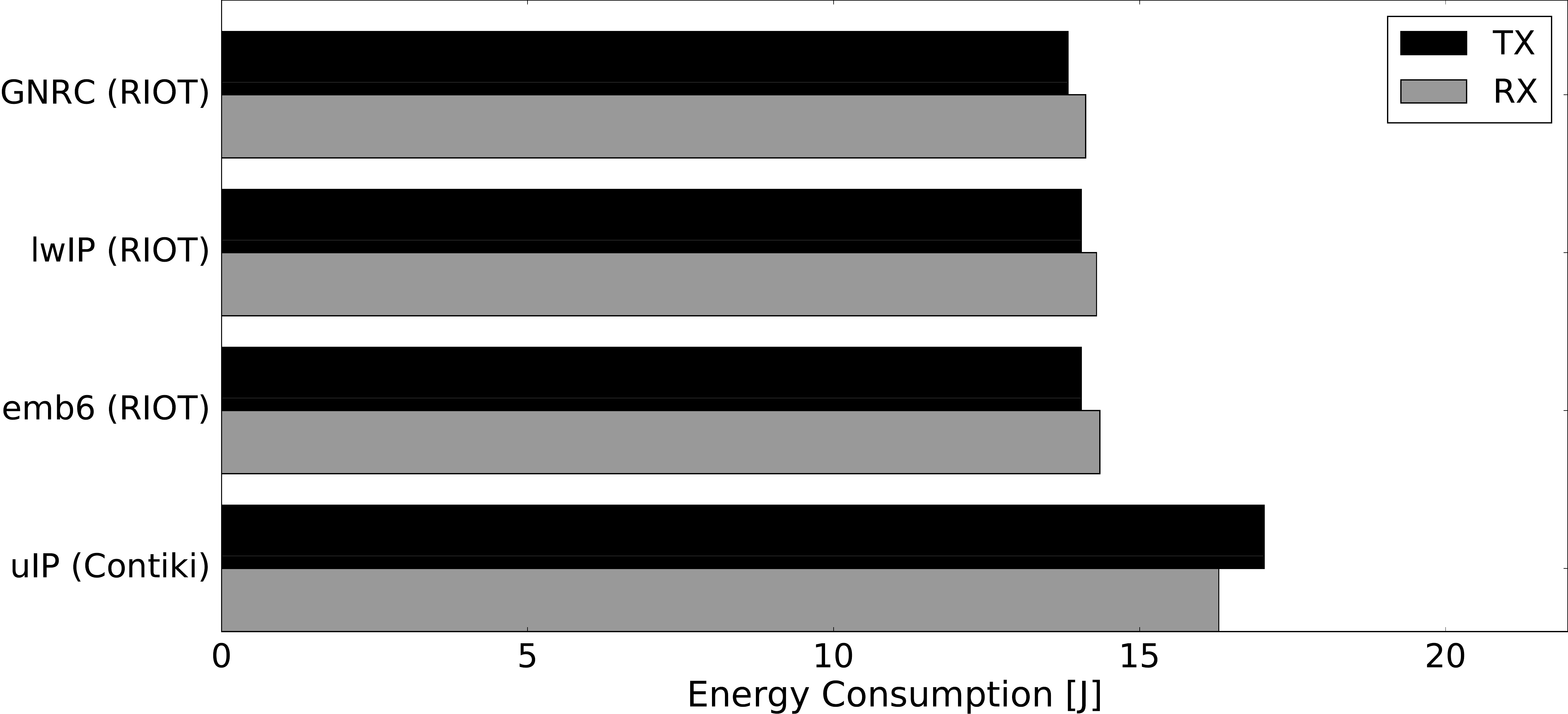}
  \caption{Energy consumption of IoT networks stacks}
  \label{fig:nw_comp_stacks_energy}
\end{figure}

\paragraph{Energy}
We now take a closer look at the total energy required for communication using different IP stacks. Figure \ref{fig:nw_comp_stacks_energy} displays the cumulative energy consumed by the sending and receiving nodes after 100 packets of 20~Bytes were transmitted at a rate of one packet per second. GNRC, lwIP, and emb6 were all run on RIOT and essentially require the same resources. Correspondingly, a RIOT node in the FIT IoT testbed spends about $140\, mJ$ per packet when communicating periodically at this rate. Such a scenario of regular workload  is largely shaped by the OS resource scheduling including sleep cycling. 
 As part of Contiki, uIP consumes about 20 \% more energy, which indicates that Contiki manages the identical hardware components in a less efficient manner.

\begin{figure}
  \centering
  \includegraphics[width=1.0\columnwidth]{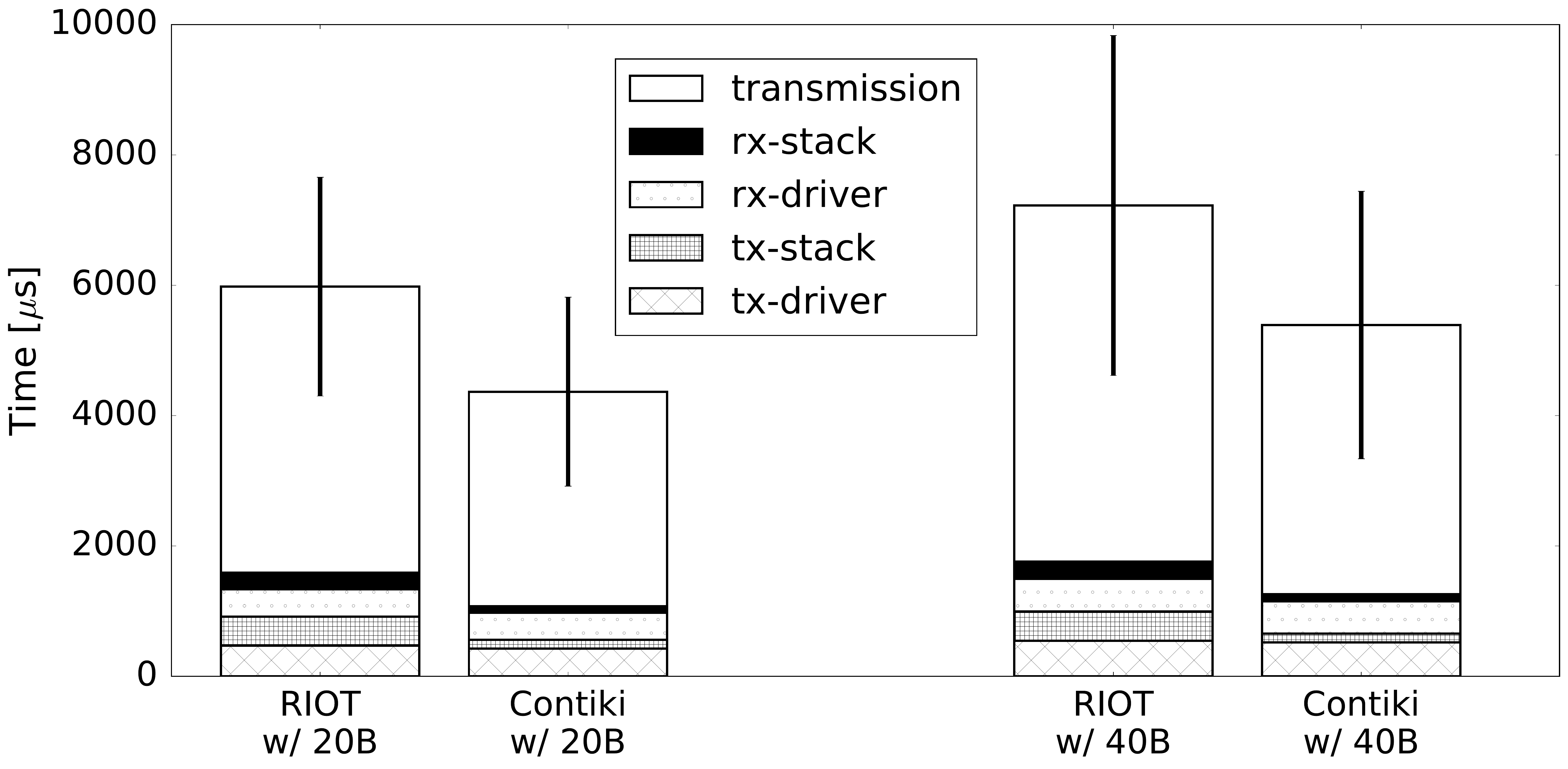}
  \caption{Device-to-device packet traveling times}
  \label{fig:on-air_contiki_riot}
\end{figure}

\paragraph{Radio Transmission: RIOT versus Contiki}
For the last examination of IP communication, we take an overall look at the node-to-node packet transmission times between RIOT (GNRC) and Contiki (uIP) systems. We probe with packets of 20 and 40 Bytes UDP payload to exclude fragmentation. Figure~\ref{fig:on-air_contiki_riot} displays the overall travel times of  packets from the sending to the receiving application with differentiation of the sending and receiving stacks, drivers, and the actual transmission by the underlying hardware. Strikingly we find the RIOT results exceeding the Contiki values by about 20 \% at very large fluctuations (see error bars). However, stack and driver processing are only slightly enhanced on RIOT, which is consistent with previous analyses. The dominating difference is generated by the hardware transmission due to a less optimized utilization of the SPI and radio buffers. Hardware utilization is also the source of the large errors. As a young OS, RIOT has still many potentials optimizing hardware support and adaptation in future work.

\subsubsection{Future Internet Stacks}

Finally, we take a somehow orthogonal perspective when looking at the
performance of an Information-centric Network stack for future Internet
communication. ICN follows a different, content-centric networking paradigm
and operates in a request-response  mode \cite{adiko-sind-12}. A request is
named Interest and carries the name of a content chunk of arbitrary length.
As a consequence, the system has higher processing overheads from parsing
and comparing these names, but the stack has fewer layers and requires less
context switches than IP. It is worth noting that the asynchronous
inter-layer-com\-mu\-nication of the RIOT networking architecture is of particular advantage to ICN. This name-based networking can apply different strategy layers for different name prefixes, which easily maps to asynchronous IPC.
In the following, we analyze a common ICN~stack, CCN-lite, which is available as a module, due to the clean RIOT networking subsystem.

\begin{figure}
  \centering
  \subfigure[ROM]{\includegraphics[width=0.48\columnwidth]{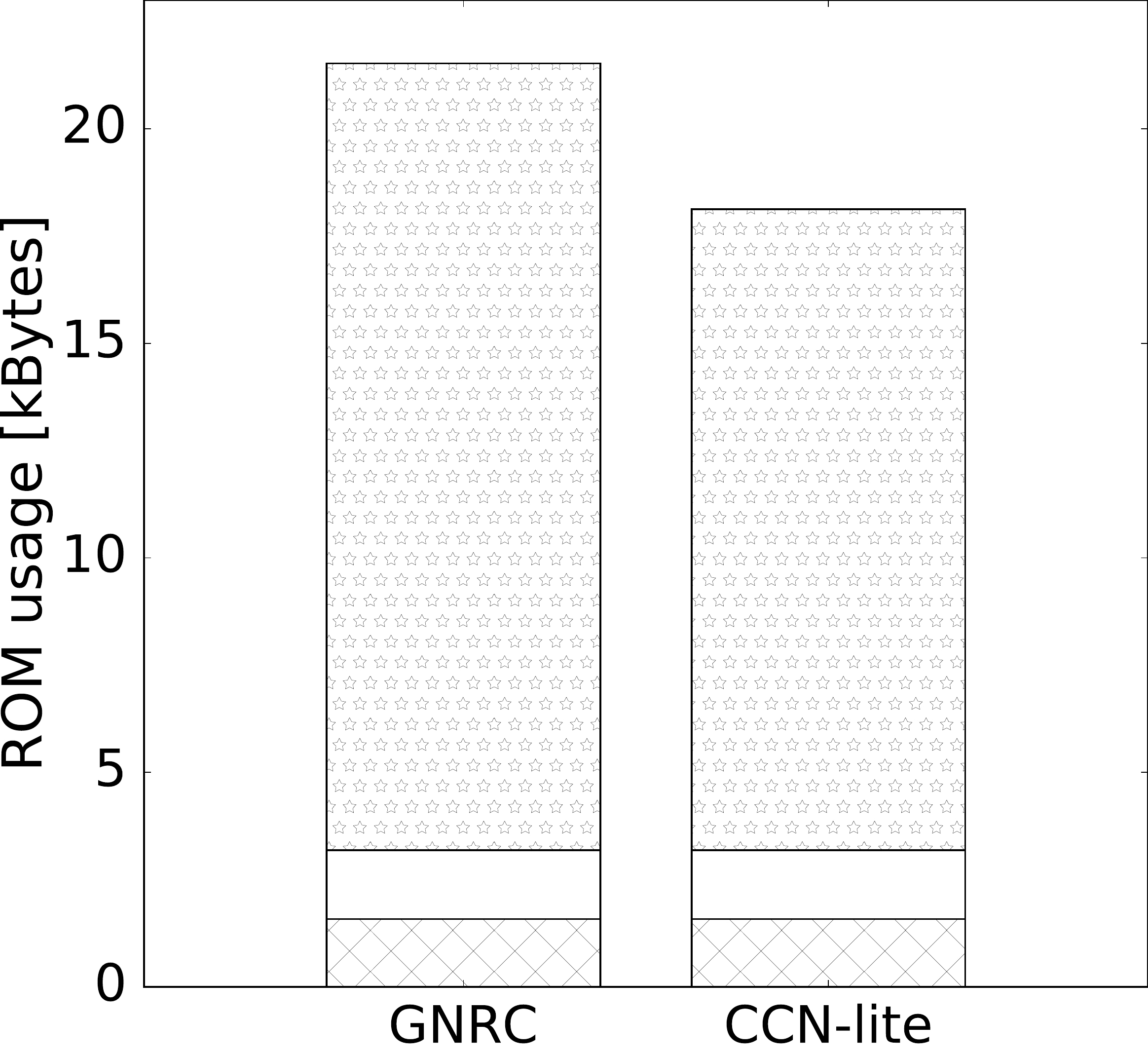}
  %\caption{Effective data transmission with 802.15.4 transceiver versus theory}
   % \label{fig:nw_comp_ram}
  }
  \subfigure[RAM]{\includegraphics[width=0.48\columnwidth]{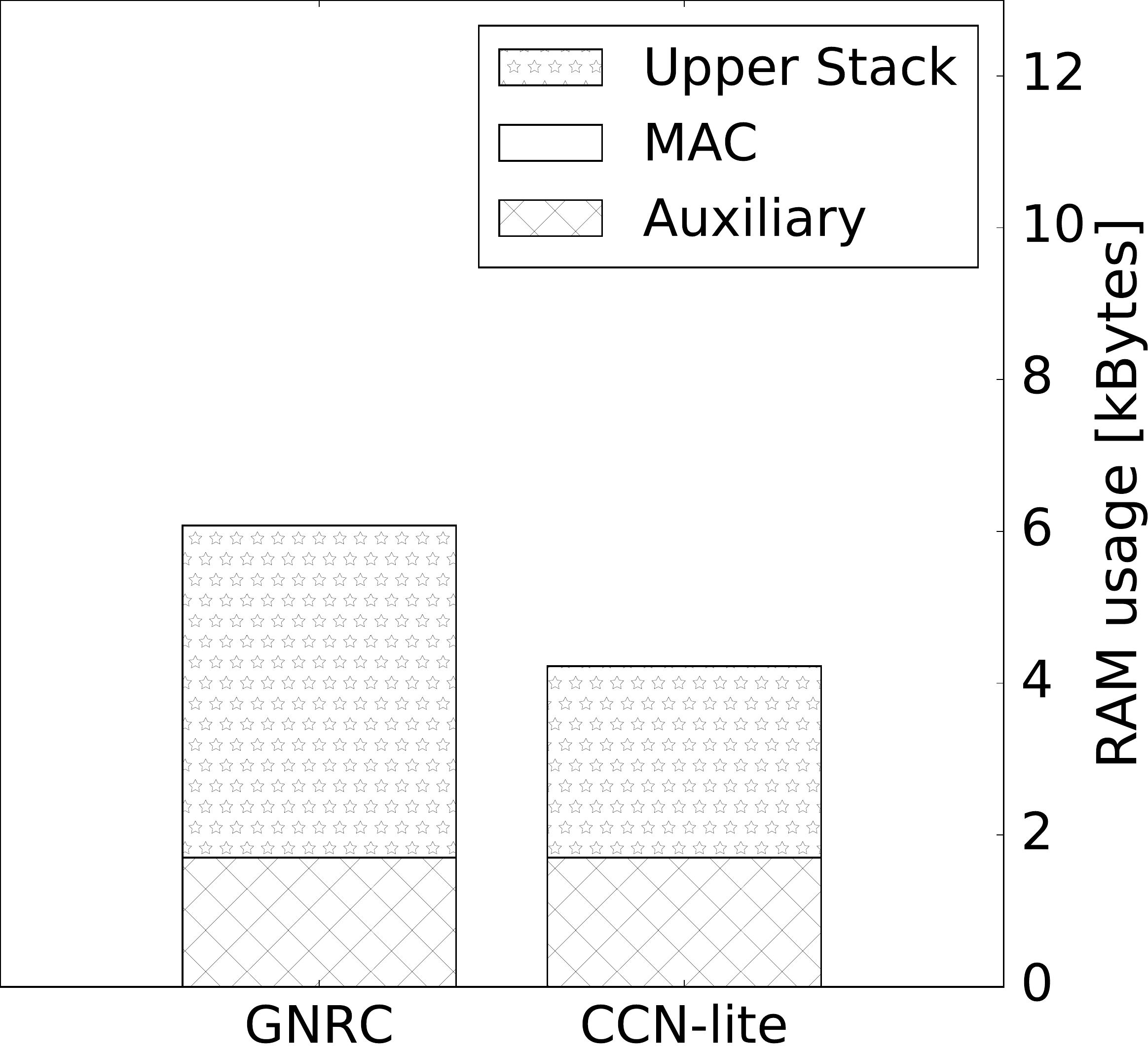}
  %\caption{Effective data reception with 802.15.4 transceiver versus theory}
  %\label{fig:nw_comp_ram}
  }
  \caption{CCN network stack size compared with GNRC}
  \label{fig:nw_ccn_rom_ram}
\end{figure}

\paragraph{Memory}
In Figure \ref{fig:nw_ccn_rom_ram}, we compare ROM and RAM of CCN-lite  with the GNRC IP stack in compatible configurations. The lower diversity and complexity of ICN protocol operations above the MAC layer is clearly visible in code size and main memory. The full-featured ICN stack consumes only 4 kB RAM, which makes this approach clearly attractive for the embedded world.

\paragraph{Processing Times}
The  processing performance of CCN-lite was measured in closest correspondence to the previous experiments on IP, but needs differentiation w.r.t. name lengths. At the sender, we measured the time between an application creating the Interest and the driver receiving the full packet for transmission. At the receiver side our measurement interval includes creating a PIT entry  with face to the application, matching a content chunk, and piping this chunk into the RX queue of the driver. Lengths of contents names were incremented up until the maximum packet size. Note that chunks carry payload, why the maximum length at the receiver is shorter. Chunk sizes are adjusted to the MTU, since ICN does not fragment.

\begin{figure}
  \centering
  \includegraphics[width=1.0\columnwidth]{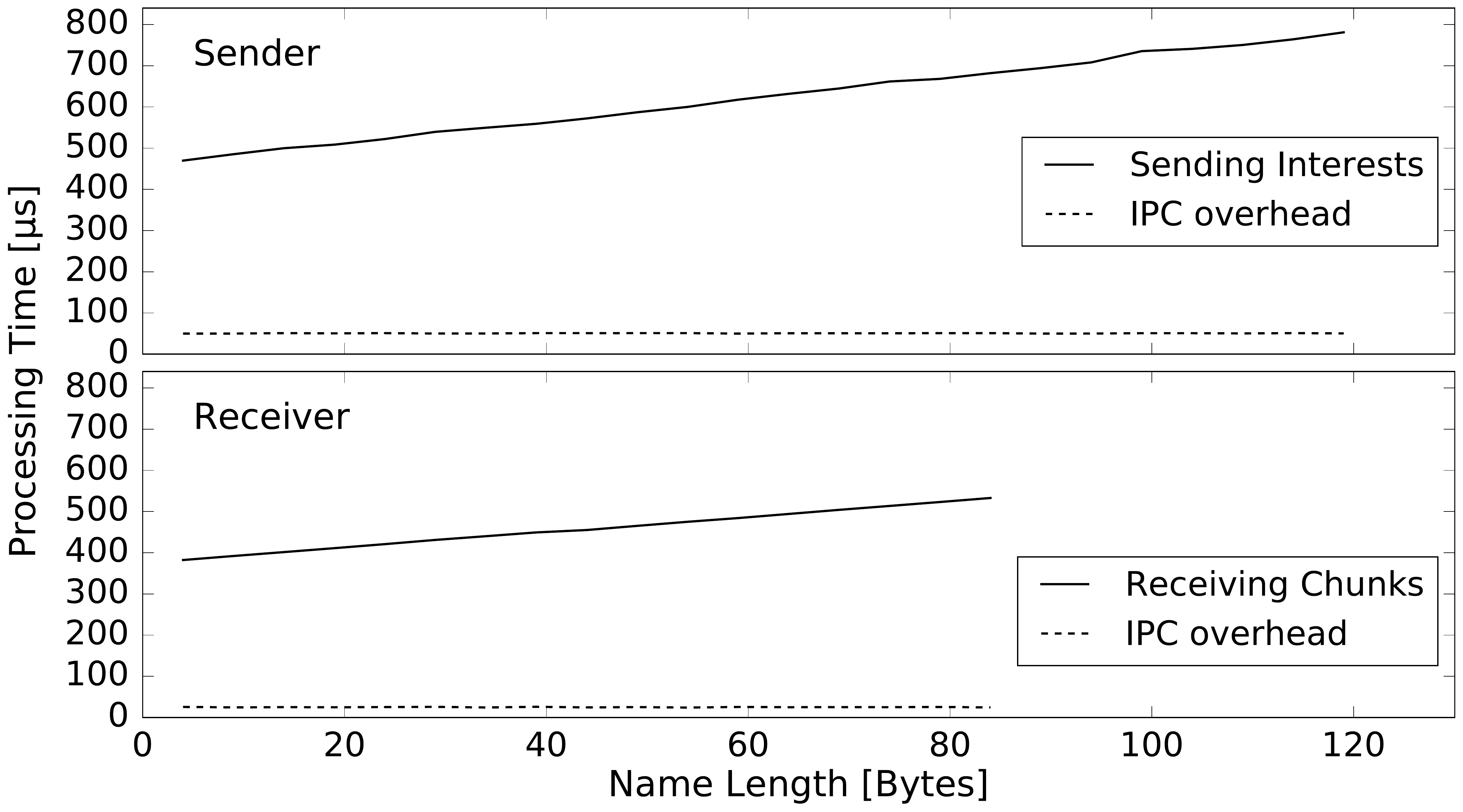}
  \caption{Processing time per request of the CCN stack}
  \label{fig:nw_ccn_proc}
\end{figure}

Figure \ref{fig:nw_ccn_proc} depicts our results for request processing times. IPC overheads are separately displayed but negligible. The CCN stack operations are significantly more costly in terms of CPU than our IP stack, cf. Figure~\ref{fig:processing_stack_all}. Even for very short names, CCN-lite consumes more than twice as many CPU cycles than GNRC, which is a clear downside for today's IoT. However, future developments are expected to contribute efficient header compression and further optimizations that will call for re-evaluations and re-consideration. 

%% file: tex/conclusion.tex
\section{Conclusions \& Future Work}
\label{sec:conclusions}

In this paper, we have introduced and analyzed the RIOT networking subsystem, a modern open source contribution for the IoT, designed and implemented from scratch.
We demonstrated that concepts of a cleanly layered, modular stack---so far only known from full-fledged operating systems---in fact do fit tiny devices.

Leveraging clean and proper interfaces, efficient data structures, and the efficient IPC of the RIOT kernel,
%we show that the IoT can be built as modern software without sacrificing performance.
we showed that such an approach enhances implementation portability and flexibility, but also maintainability over time. 
Our thorough, comparative evaluations of various metrics for several network stacks and their micro-components show that these gains were not at the expense of performance.

In the future, we will work on the integration of additional upcoming standards (e.g., IP over lpWAN).
We will also work on leveraging multi-interface support in our stack to enable adaptive multi-path forwarding.
To increase robustness and faster data delivery even further, we will explore in-network caching in more detail.

%Typically, an operating system is a complex piece of software that is always work-in-progress (e.g., see Linux).
%Future work includes, but is not limited to, {\bf network stack evolution} to integrate both upcoming standards (e.g. IP over lpWAN) and novel research concepts (e.g. ICN for IoT), {\bf system security evolution} to provide partial or full IoT software updates and to leverage upcoming privileged modes (e.g. TrustZone) on mid-range IoT hardware, {\bf community processes evolution} to adapt to an ever-growing RIOT community, keeping IETF and Linux communities as guiding inspiration.

\iffalse
An OS (even and particular a stripped-down OS such as this one) is a complex piece of software that is always work-in-progress (see Linux).

\paragraph{Network stack future work} 
incorporate new standard technologies and protocols (gnrc is designed to facilitate such evolution). Give examples.

\paragraph{Security} 
Need to somehow say that RIOT as it is now is a good base to provide not only network security, but also system security aspects. 

\paragraph{Development processes future work} 
test framework designed to evolve naturally with relevant hardware as it becomes available. Give other examples.

\paragraph{Community processes evolution} 
Challenge: scaling open processes. Give example of some open process which has evolved, or is evolving. Highlight IETF and Linux community as our models for such things.
\fi